\documentstyle[12pt]{article}
\def\simlt{\mathrel{\lower2.5pt\vbox{\lineskip=0pt\baselineskip=0pt
           \hbox{$<$}\hbox{$\sim$}}}}
\def\simgt{\mathrel{\lower2.5pt\vbox{\lineskip=0pt\baselineskip=0pt
           \hbox{$>$}\hbox{$\sim$}}}}
\def\msq{m_{\tilde{q}}}
\def\msta{m_{\tilde{t}_1}}
\def\mstb{m_{\tilde{t}_2}}
\def\msba{m_{\tilde{b}_1}}
\def\msbb{m_{\tilde{b}_2}}

\def\mch{m_{H^{\pm}}}

\def\mev{\; {\rm MeV}}
\def\gev{\; {\rm GeV}}
\def\tev{\; {\rm TeV}}
\def\pb{\; {\rm pb}}
\def\to{\rightarrow}
\def\ov{\overline}
\def\tb{\tan \beta}

\def\sabsq{\sin^2 (\beta - \alpha)}
\def\cabsq{\cos^2 (\beta - \alpha)}
\newcommand{\be}{\begin{equation}}
\newcommand{\ee}{\end{equation}}
\def\ln {{\rm ln}}

\def\abs#1{\left| #1\right|}
\def\lapp{{\ \lower 0.6ex \hbox{$\buildrel<\over\sim$}\ }}
\def\gapp{{\ \lower 0.6ex \hbox{$\buildrel>\over\sim$}\ }}
\def\to{\rightarrow}
\def\as{\alpha_s}

\def\pp{{\rm p\bar{p}}}
\def\bb{b\bar{b}}
\def\tt{t\bar{t}}

\def\pb{{\rm pb}}
\def\fb{{\rm fb}}
\def\s2tw{{\sin}^2\theta_W}

\def\epem{\ifmmode e^+e^-\else $e^+e^-$ \fi}
\def\taptam{\ifmmode \tau^+\tau^-\else $\tau^+\tau^-$ \fi}
\def\mupmum{\ifmmode \mu^+\mu^-\else $\mu^+\mu^-$ \fi}
\def\pp{\ifmmode p\bar{p} \else $p\bar{p}$ \fi}
\def\bb{\ifmmode b\bar{b} \else $b\bar{b}$ \fi}
\def\tt{\ifmmode t\bar{t} \else $t\bar{t}$ \fi}
\def\ep{\ifmmode e^-p \else $e^-p$ \fi}
\def\MSB{\ifmmode{\overline{\rm MS}}\else $\overline{\rm MS}$\ \fi}

\def\as{\ifmmode \alpha_S \else $\alpha_S$ \fi}
\def\nf{\ifmmode \n_{\rm f} \else $\n_{\rm f}$ \fi}

\newcommand{\bea}{\begin{eqnarray}}
\newcommand{\eea}{\end{eqnarray}}
\newcommand{\bean}{\begin{eqnarray*}}
\newcommand{\eean}{\end{eqnarray*}}

\newcommand{\gapproxeq}{\lower .7ex\hbox{$\;\stackrel{\textstyle >}{\sim}\;$}}
\newcommand{\lapproxeq}{\lower .7ex\hbox{$\;\stackrel{\textstyle <}{\sim}\;$}}

\begin{document}
\begin{titlepage}
\vspace*{-1cm}
\noindent
\phantom{DRAFT}
\hfill{CERN-TH.6150/91}
\newline
\phantom{December 1991}
\hfill{ETH-TH/91-7}
\vskip 2.5cm
\begin{center}
{\Large\bf Testing the Higgs Sector of the}
\end{center}
\begin{center}
{\Large\bf Minimal Supersymmetric Standard Model}
\end{center}
\begin{center}
{\Large\bf at Large Hadron Colliders }
\end{center}
\vskip 1.5cm
\begin{center}
{\large Z. Kunszt}  \\
Institute of Theoretical Physics, ETH, \\
Zurich, Switzerland  \\
\vskip .3cm
and \\
\vskip .3cm
{\large F. Zwirner}\footnote{On leave from INFN, Sezione di Padova, Italy.}
\\
Theory Division, CERN, \\
Geneva, Switzerland \\
\vskip 1cm
\end{center}
\begin{abstract}
\noindent
We study the Higgs sector of the Minimal Supersymmetric Standard Model,
in the context of proton-proton collisions at LHC and SSC energies.
We assume a relatively heavy
supersymmetric particle spectrum, and include recent results on
one-loop radiative corrections to Higgs-boson masses and couplings.
We begin by discussing present and future constraints from the LEP
experiments. We then compute branching ratios and total widths
for the neutral ($h,H,A$) and charged ($H^\pm$) Higgs particles.
We present total cross-sections and event rates for the important
discovery channels at the LHC and SSC. Promising physics signatures
are given by $h \to \gamma \gamma$, $H \to \gamma \gamma$ or $Z^* Z^*$
or $\tau^+ \tau^-$, $A \to \tau^+ \tau^-$, and $t \to b H^+$ followed
by $H^+ \to \tau^+ \nu_{\tau}$, which should allow for an almost complete
coverage of the parameter space of the model.
\end{abstract}
\vfill{
CERN-TH.6150/91
\newline
\noindent
ETH-TH/91-7
\newline
\noindent
December 1991}
\end{titlepage}
\vskip2truecm
\section{Introduction}
All available experimental data in particle physics are consistent with
the Standard Model (SM) of strong and electroweak interactions, provided
[\ref{LPHEP91}]
\be
\label{mtop}
91 \gev < m_t < 180 \gev \;\;\; (95 \% \, {\rm c.l.})
\ee
and
\be
\label{mhiggs}
57 \gev < m_{\varphi} \;\;\; (95 \% \, {\rm c.l.}) \, ,
\ee
where $m_t$ and $m_{\varphi}$ denote the masses of the top quark
and of the SM Higgs boson, respectively. The lower limits on
$m_t$ and $m_{\varphi}$ are obtained from unsuccessful direct
searches at the Tevatron and LEP. The upper limit on $m_t$ is
obtained as a consistency condition of the SM, after the inclusion of
radiative corrections, with the high-precision data on electroweak
phenomena. Strong evidence for the existence of the top quark, with
the quantum numbers predicted by the SM, is also provided by the
precise measurements of the weak isospin of the $b$-quark.
In the case of the Higgs boson, the situation is radically different.
There is no experimental evidence yet that the minimal SM Higgs
mechanism is the correct description of electroweak symmetry breaking.
Fortunately, present and future accelerators will give decisive
contributions towards the experimental solution of this problem.
If the SM description of the Higgs mechanism is correct, LEP or the LHC
and SSC should be able to find the SM Higgs boson and study its
properties.

Despite its remarkable successes, the SM can only be regarded as an
effective low-energy theory, valid up to some energy scale $\Lambda$
at which it is replaced by some more fundamental theory. Certainly
$\Lambda$ is less than the Planck scale, $M_{\rm P} \sim 10^{19} \gev$, since
one needs a theory of quantum gravity to describe physics at these energies.
However, there are also arguments, originating precisely from the study of
the untested Higgs sector \footnote{For reviews of Higgs boson physics see,
e.g., refs. [\ref{higgs},\ref{hunter}]}, which suggest that $\Lambda$ should
rather be close to the Fermi scale $G_{\rm F}^{-1/2} \sim 300 \gev$. The
essence of these arguments is the following. Triviality of the $\lambda
\varphi^4$ theory, absence of Landau poles and perturbative unitarity imply
that within the SM $m_{\varphi} <$ 600--800 GeV. If one then tries to extend
the validity of the SM to energy scales $\Lambda \gg G_{\rm F}^{-1/2}$, one is
faced with the fact that in the SM there is no symmetry to justify the
smallness of the Higgs mass with respect to the (physical) cut-off
$\Lambda$. This is apparent from the fact that in the SM one-loop
radiative corrections to the Higgs mass are quadratically divergent;
it is known as the naturalness (or hierarchy) problem of the SM.
Motivated by this problem, much theoretical effort has been devoted
to finding descriptions of electroweak symmetry breaking which modify the
SM at scales $\Lambda \sim G_{\rm F}^{-1/2}$. The likely possibility of
such modifications is the reason why, when discussing the experimental
study of electroweak symmetry breaking, one should not be confined to
the SM Higgs, but also consider alternatives to it, which might have
radically different
signatures, and in some cases be more difficult to detect than the SM Higgs.
Only after a thorough study of these alternatives can one be
definite about the validity of the so-called `no-lose theorems', stating that
the physics signatures of electroweak symmetry breaking cannot be missed
at LEP or the LHC and SSC.

When considering alternatives to the minimal SM Higgs sector, it is natural to
concentrate on models which are theoretically motivated, phenomenologically
acceptable and calculationally well-defined. The most attractive possibility
satisfying these criteria is the Minimal Supersymmetric Standard Model (MSSM)
[\ref{MSSM}].
This possibility is theoretically motivated by the fact that low-energy
supersymmetry, effectively broken in the vicinity of the electroweak scale,
is the only theoretical framework that can {\em naturally} accommodate
{\em elementary} Higgs bosons. The simplest and most predictive realization of
low-energy supersymmetry is the MSSM, defined by 1)~minimal gauge group:
$SU(3)_C \times
SU(2)_L \times U(1)_Y$; 2) minimal particle content: three generations of
quarks and leptons and two Higgs doublets, plus their superpartners; 3) an
exact discrete $R$-parity, which guarantees (perturbative) baryon- and
lepton-number conservation: $R=+1$ for SM particles and Higgs bosons,
$R=-1$ for their superpartners; 4) supersymmetry breaking parametrized by
explicit but soft breaking terms: gaugino and scalar masses and trilinear
scalar couplings.

Besides the solution of the naturalness problem, there are other virtues of
the MSSM which are not shared by many other alternatives to the SM Higgs and
should also be recalled to further motivate our study. The MSSM successfully
survives all the stringent phenomenological tests coming from precision
measurements at LEP: in most of its parameter space, the MSSM predictions
for the LEP observables are extremely close to the SM predictions, evaluated
for a relatively light SM Higgs [\ref{susyrad}]. This can be compared, for
example, with the simplest technicolor models, which are ruled out by the
recent LEP data [\ref{guido}]. Again in contrast with models of dynamical
electroweak symmetry breaking,
the MSSM has a high degree of predictivity, since all masses and couplings
of the Higgs boson sector can be computed, at the tree-level, in terms of
only two parameters, and radiative corrections can be kept under control:
in particular, cross-sections and branching ratios for the MSSM Higgs bosons
can be reliably computed in perturbation theory. Furthermore, it is
intriguing that the idea of grand unification, which fails in its minimal
non-supersymmetric implementation, can be successfully combined with
that of low-energy supersymmetry: minimal supersymmetric grand unification
predicts a value of $\sin^2 \theta_W (m_Z)$ which is in good agreement with
the measured one, and a value of the grand-unification mass which
could explain why proton decay has escaped detection
so far [\ref{ross}]. Finally, as a consequence of $R$-parity, the lightest
supersymmetric
particle, which is typically neutral and weakly interacting, is absolutely
stable, and thus a natural candidate for dark matter.

Any consistent supersymmetric extension of the SM requires at least two Higgs
doublets, in order to give masses to all charged quarks and leptons and to
avoid gauge anomalies originated by the spin-$1/2$ higgsinos. The MSSM
has just two complex Higgs doublets, with the following $SU(3)_C \times
SU(2)_L \times U(1)_Y$ quantum numbers ($Q=T_{3L}+Y$):
\be
H_1 \equiv
\pmatrix
{H_1^0 \cr H_1^- \cr}
\sim ( 1, 2, -1/2) \, ,
\;\;\;\;\;
H_2 \equiv
\pmatrix
{H_2^+ \cr H_2^0}
\sim ( 1, 2, +1/2) \, .
\ee
Other non-minimal models can be constructed, but they typically increase
the number of parameters without correspondingly increasing the physical
motivation. For example, the simplest non-minimal model [\ref{nmssm}]
is constructed
by adding a singlet Higgs field $N$ and by requiring purely trilinear
superpotential couplings. In this model, the Higgs sector has already
two more parameters than in the MSSM. Folklore arguments in favour of
this model are that it avoids the introduction of a supersymmetry-preserving
mass parameter $\mu \sim G_{\rm F}^{-1/2}$ and that the homogeneity properties
of its superpotential recall the structure of some superstring effective
theories. A closer look, however, shows that these statements should be
taken with a grain of salt. First, in the low-energy effective theory
with softly broken global supersymmetry, the supersymmetric mass $\mu \sim
G_{\rm F}^{-1/2}$ could well be a remnant of local supersymmetry breaking,
if the underlying supergravity theory has a suitable structure of
interactions [\ref{muproblem}]. Moreover, when embedded in a grand-unified
theory, the non-minimal model with a singlet Higgs field might develop
dangerous instabilities [\ref{singlet}]. Also, the trilinear $N^3$
superpotential coupling, which is usually invoked to avoid a massless axion,
is typically absent in string models. We therefore concentrate in this paper
on the MSSM only.

The previous considerations should have convinced the reader that the Higgs
sector of the MSSM is worth a systematic study in view of the forthcoming
hadron colliders, the LHC and SSC. To perform such a study, one has to
deal with the rich particle spectrum of the MSSM. As discussed in more detail
later, the Higgs sector contains one charged ($H^\pm$) and three neutral
($h,H,A$) physical states. At the classical level, all
Higgs boson masses and couplings can be expressed in terms of
two parameters only, for example $m_A$ and $\tan \beta \equiv
v_2/v_1$. This makes the discussion more complicated than in the
SM, where the only free parameter in the Higgs sector is the Higgs
mass, $m_{\varphi}$.
In addition, when considering production and decay of Higgs
bosons, the whole particle spectrum of the model has to be
taken into account. As in the SM, the top-quark mass $m_t$
is an important parameter: barring the fine-tuned cases of a very light
stop squark, or of charginos very close in mass to $m_Z/2$,
the limits of eq. (1) are also valid in the MSSM [\ref{susyrad}].
In contrast with the SM, also the supersymmetric $R$-odd particles
(squarks, sleptons, gauginos, higgsinos) can play an important role in
the production and decay of supersymmetric Higgs bosons [\ref{gunion}].
Clearly, to keep track simultaneously of all supersymmetric-particle masses
would be a difficult (and confusing) task. We shall therefore
concentrate, following the approach of ref. [\ref{kz}], on the limiting
case where all supersymmetric-particle masses
are heavy enough not to play an important role in the phenomenology
of supersymmetric Higgs bosons. This is phenomenologically meaningful,
since one can argue that a relatively light
supersymmetric-particle spectrum is likely to give independent,
detectable signatures at LEP or at the LHC and SSC.

Another motivation for the present study is the recent realization
[\ref{pioneer}] that
tree-level formulae for Higgs-boson masses and couplings can receive large
radiative corrections, dominated by the exchange of virtual top and bottom
quarks and squarks in loop diagrams. For example, tree-level formulae would
predict the existence of a neutral Higgs boson ($h$) lighter than the $Z$.
If this were true, there would be a chance of testing completely
the MSSM Higgs sector at LEP~II, with no need for the LHC and SSC.
However, $m_h$ can receive a large positive shift by radiative
corrections, which can push $h$ beyond the LEP II discovery reach.
This makes the LHC and SSC important, not only for a possible confirmation
of a SUSY Higgs signal seen at LEP, but also for the exploration of
the parameter space inaccessible to LEP.

The phenomenology of the SM Higgs at the LHC
[\ref{zkwjsaa}--\ref{evian}]
and SSC [\ref{snow},\ref{sdcgem}] has been intensely
studied over the last years: a lot of effort was required to prove
[\ref{zkwjsaa},\ref{froid}], at
least on paper, that the combination of LEP and the LHC/SSC is sufficient to
explore the full theoretically allowed range of SM Higgs masses.
However, those results cannot be directly applied to the neutral states of
the MSSM, since there are important differences in the couplings, and of
course one needs to analyse separately the case of the charged Higgs.
Even in the case in which all the $R$-odd supersymmetric particles are very
heavy, the Higgs sector of the MSSM represents a non-trivial extension of
the SM case. Also several studies of the MSSM Higgs sector have already
appeared in the literature. In particular, tree-level formulae for
the MSSM Higgs boson masses and couplings are available, and they have
already been used to compute cross-sections and branching ratios for
representative values of the MSSM parameters [\ref{hunter}].
However, the existing
analyses are not systematic enough to allow for a definite conclusion
concerning the discovery potential of the LHC and SSC, even in the simple
case of large sparticle masses. Also, they do not include radiative
corrections to Higgs-boson masses and couplings. In this paper we plan
to help filling these two gaps.
The strategy for a systematic study of neutral supersymmetric Higgs
bosons at the LHC was outlined in ref. [\ref{kz}]: however, at that time
radiative corrections were not available, and also the $\gamma \gamma$
branching ratio was incorrectly encoded in the computer program.
Our goal will be to see if LEP and the LHC/SSC can be sensitive
to supersymmetric Higgs bosons in the whole $(m_A,\tan \beta)$ space.

The structure of the paper is the following. In sect.~2
we review the theoretical structure of the Higgs sector
of the MSSM, including radiatively corrected formulae for
Higgs-boson masses and couplings. In sect.~3 we survey
the present LEP I limits, after the inclusion of radiative corrections,
and the plausible sensitivity of LEP II. In sect.~4
we present branching ratios and widths of neutral and
charged supersymmetric Higgs bosons. In sect.~5
we compute the relevant cross-sections at the LHC and
SSC, and in sect.~6 we examine in some detail the most
promising signals for
discovery. Finally, sect.~7 contains a concluding discussion
of our results and of prospects for further work.
\section{Higgs masses and couplings in the MSSM}
For a discussion of Higgs-boson masses and couplings in the MSSM, the
obvious starting point is the tree-level Higgs potential [\ref{MSSM}]
\bea
\label{V0}
V_0 &  = &  m_1^2 \left| H_1 \right|^2 + m_2^2 \left| H_2
\right|^2 + m_3^2 \left( H_1 H_2 + {\rm h.c.} \right) \nonumber \\
& &
+ {1 \over 8}g^2 \left( H_2^{\dagger} {\vec\sigma} H_2
+ H_1^{\dagger} {\vec\sigma} H_1 \right)^2 +
{1 \over 8}g'^2 \left(  \left| H_2 \right|^2 -
\left| H_1 \right|^2 \right)^2,
\eea
where $m_1^2,m_2^2,m_3^2$ are essentially arbitrary mass parameters,
$g$ and $g'$ are the $SU(2)$ and $U(1)$ coupling constants,
respectively, and ${\vec\sigma}$ are the Pauli matrices. $SU(2)$
indices are left implicit and contracted in the obvious way. It is not
restrictive to choose $m_3^2$ real and negative, and then the vacuum
expectation values $v_1 \equiv \langle H_1^0 \rangle$ and $v_2 \equiv
\langle H_2^0 \rangle$ real and positive.

The physical states of the MSSM Higgs sector are three neutral bosons
(two CP-even, $h$ and $H$, and one CP-odd, $A$) and a charged boson,
$H^{\pm}$. A physical constraint comes from the fact that the combination
$(v_1^2+v_2^2)$, which determines the $W$ and $Z$ boson masses, must
reproduce their measured values. Once this constraint is imposed, in the Born
approximation the MSSM Higgs sector contains only two independent parameters.
A convenient choice, which will be adopted throughout this paper, is to take as
independent parameters $m_A$, the physical mass of the CP-odd neutral boson,
and $\tan \beta \equiv v_2 / v_1$, where $v_1$ gives mass to charged leptons
and quarks of charge $-1/3$, $v_2$ gives mass to quarks of charge $2/3$.
The parameter $m_A$ is essentially unconstrained, even if naturalness
arguments suggest that it should be smaller than $O(500 \gev)$, whereas
for $\tan \beta$ the range permitted by model calculations
is $1 \le \tan \beta \simlt \frac{m_t}{m_b}$.

At the classical level, the mass matrix of neutral CP-even Higgs bosons
reads
\be
\label{cpeven0}
\left( {\cal M}_R^0 \right)^2 =
\left[
\pmatrix{
\cot\beta & -1 \cr
-1 & \tan\beta \cr}
{m_Z^2 \over 2}
+
\pmatrix{
\tan\beta & -1 \cr
-1 & \cot\beta \cr}
{m_A^2 \over 2}
\right]
\sin 2\beta
\ee
and the charged-Higgs mass is given by
\be
\label{mch0}
m_{H^\pm}^2 = m_W^2 + m_A^2.
\ee
{}From eq. (\ref{cpeven0}), one obtains
\be
\label{mh0}
m_{h,H}^2 = {1 \over 2} \left[
m_A^2 + m_Z^2 \mp \sqrt{(m_A^2 + m_Z^2)^2
- 4 m_A^2 m_Z^2 \cos^2 2 \beta}
\right],
\ee
and also celebrated inequalities such as $m_W, m_A < m_{H^{\pm}}$,
$m_h <  m_Z < m_H$, $m_h <  m_A < m_H$. Similarly, one can easily compute all
the Higgs-boson couplings by observing that the mixing angle $\alpha$,
required to diagonalize the mass matrix (\ref{cpeven0}), is given by
\be
\cos 2 \alpha = - \cos 2 \beta \; { { m_A^2  - m_Z^2 }\over {
 m_H^2 - m_h^2  }} \ ,\ \ - {\pi \over 2} < \alpha\  {\leq}\  0.
\ee
For example, the couplings of the three neutral Higgs bosons
are easily obtained from the SM Higgs couplings if
one multiplies them by the $\alpha$- and $\beta$-dependent factors
summarized in table~1.
\begin{table}
\begin{center}
\begin{tabular}{|c|c|c|c|}
\hline
& & & \\
&
$
\begin{array}{c}
d\overline{d},s\overline{s},b\overline{b} \\
e^+e^-,\mu^+\mu^-,\tau^+\tau^-
\end{array}
$
& $u\overline{u},c\overline{c},t\overline{t}$
& $W^+W^-,ZZ$ \\
& & & \\
\hline
& & & \\
$h$
& $- \sin \alpha / \cos \beta $
& $  \cos \alpha / \sin \beta $
& $  \sin \, (\beta -\alpha)  $ \\
& & & \\
\hline
& & & \\
$H$
& $\cos \alpha / \cos \beta$
& $\sin \alpha / \sin \beta$
& $\cos \, (\beta -\alpha)  $
\\
& & & \\
\hline
& & & \\
$A$
& $-i\gamma_5 \tan \beta $
& $-i\gamma_5 \cot \beta $
& $0$
\\
& & & \\
\hline
\end{tabular}
\caption{Correction factors for the couplings of the MSSM neutral Higgs
bosons to fermion and vector boson pairs.}
\label{couplings}
\end{center}
\end{table}
The remaining tree-level Higgs-boson couplings in the MSSM can be easily
computed and are summarized, for example, in ref. [\ref{hunter}]. An
important consequence of the structure of the classical Higgs potential
of eq. (\ref{V0}) is the existence of at least one neutral CP-even Higgs
boson, weighing less than or about $m_Z$ and with approximately standard
couplings to the $Z$. This raised the hope that the crucial experiment
on the MSSM Higgs sector could be entirely performed at LEP II (with
sufficient centre-of-mass energy, luminosity and $b$-tagging efficiency),
and took some interest away from the large hadron collider environment.

However, it was recently pointed out [\ref{pioneer}] that
the masses of the Higgs bosons in the MSSM are subject to large radiative
corrections, associated with the top quark and its $SU(2)$ and supersymmetric
partners\footnote{Previous studies [\ref{previous}] either
neglected the case of a heavy top quark, or
concentrated on the violations of the neutral-Higgs mass sum rule
without computing corrections to individual Higgs masses.}.
Several papers [\ref{higgscorr}--\ref{berz}] have subsequently investigated
various aspects of these corrections and their implications for experimental
searches at LEP. In the rest of this section, we shall summarize and
illustrate the main effects of radiative corrections on Higgs-boson
parameters.

As far as Higgs-boson masses and self-couplings are concerned, a convenient
approximate way of parametrizing one-loop radiative corrections is to
substitute the tree-level Higgs potential of eq. (\ref{V0}) with the one-loop
effective potential, and to identify Higgs-boson masses and self-couplings
with the appropriate combinations of derivatives of the effective potential,
evaluated at the minimum.
The comparison with explicit diagrammatic calculations
shows that the effective potential approximation is more than adequate for
our purposes. Also, inspection shows that the most important corrections
are due to loops of top and bottom quarks and squarks.
At the minimum  $ \langle H_1^0 \rangle = v_1$,
$\langle  H_2^0 \rangle = v_2$, $ \langle H_1^- \rangle = \langle H_2^+
\rangle = 0$, and neglecting intergenerational mixing, one obtains for
the top and bottom quark and squark masses the familiar expressions
\be
\label{mq}
m_t^2 = h_t^2 v_2^2 \, ,
\;\;\;\;\;
m_b^2 = h_b^2 v_1^2 \, ,
\ee
\begin{eqnarray}
\label{stopeigen}
m_{\tilde{t}_{1,2}}^2 &=& m_t^2+\frac{1}{2}(m_Q^2+m_U^2)+
\frac{1}{4}m_Z^2\cos 2\beta
\phantom{aaaaaaaaaaaaaaaaaaaaaaaaaaaaaaaaaaaa}\nonumber \\
& & \pm
\sqrt{\left[\frac{1}{2}(m_Q^2-m_U^2)
+\frac{1}{12}(8m_W^2-5m_Z^2)\cos 2\beta\right]^2
+m_t^2\left(A_t+\mu\cot\beta\right)^2},\nonumber \\
\hfill
\end{eqnarray}
\begin{eqnarray}
\label{sbottomeigen}
m_{\tilde{b}_{1,2}}^2 &=& m_b^2+\frac{1}{2}(m_Q^2+m_D^2)-
\frac{1}{4}m_Z^2\cos 2\beta
\phantom{aaaaaaaaaaaaaaaaaaaaaaaaaaaaaaaaaaaa}\nonumber \\
& & \pm
\sqrt{\left[\frac{1}{2}(m_Q^2-m_D^2)
-\frac{1}{12}(4m_W^2-m_Z^2)\cos 2\beta\right]^2
+m_b^2\left(A_b+\mu\tan\beta\right)^2}.\nonumber \\
\hfill
\end{eqnarray}
In eqs. (\ref{mq}) to (\ref{sbottomeigen}), $h_t$ and $h_b$ are the top and
bottom Yukawa couplings, and $m_Q,m_U,m_D$ are soft supersymmetry-breaking
squark masses. The parameters $A_t$, $A_b$ and $\mu$, which determine the
amount of mixing in the stop and sbottom mass matrices, are defined
by the trilinear potential terms
$h_t A_t ( \tilde{t}_L \tilde{t^c}_L H_2^0
- \tilde{b}_L \tilde{t^c}_L H_2^+ ) + \, {\rm h.c.}$,
$h_b A_b ( \tilde{b}_L \tilde{b^c}_L H_1^0 -
\tilde{t}_L \tilde{b^c}_L H_1^-) + \, {\rm h.c.}$
and by the superpotential mass term $\mu (H_1^0 H_2^0 - H_1^- H_2^+)$,
respectively.

To simplify the discussion, in the following we will take a
universal soft supersymmetry-breaking squark mass,
\be
\label{soft}
m_Q^2 = m_U^2 = m_D^2 \equiv \msq^2\, ,
\ee
and we will assume negligible mixing in the stop and sbottom mass
matrices,
\be
\label{mixing}
A_t = A_b = \mu = 0 \, .
\ee
Formulae valid for arbitrary values of the parameters can be found in
refs. [\ref{erz3},\ref{berz}], but the qualitative features corresponding
to the parameter choice of eqs. (\ref{soft}) and (\ref{mixing}) are
representative of a very large region of parameter space. In the case
under consideration, and neglecting $D$-term contributions to the
field-dependent stop and sbottom masses, the neutral CP-even mass matrix
is modified at one loop as follows
\be
\label{cpeven1}
{\cal M}_R^2 = \left( {\cal M}_R^0 \right)^2
+
\pmatrix{
\Delta_1^2 & 0 \cr
0 & \Delta_2^2 \cr},
\ee
where
\be
\label{delta1}
\Delta_1^2 = {3 g^2 m_b^4 \over 16 \pi^2 m_W^2 \cos^2\beta}
\log\frac{m_{\tilde{b}_1}^2 m_{\tilde{b}_2}^2}{m_b^4},
\ee
\be
\label{delta2}
\Delta_2^2= {3 g^2 m_t^4 \over 16 \pi^2 m_W^2 \sin^2\beta}
\log\frac{m_{\tilde{t}_1}^2 m_{\tilde{t}_2}^2}{m_t^4}.
\ee
{}From the above expressions one can easily derive the one-loop-corrected
eigenvalues $m_h$ and $m_H$, as well as the mixing angle $\alpha$ associated
with the one-loop-corrected mass matrix (\ref{cpeven1}).
The one-loop-corrected charged Higgs mass is given instead by
\begin{equation}
\label{sumrule}
m_{H^{\pm}}^2=
m_W^2 + m_A^2 +
\Delta^2,
\end{equation}
where, including $D$-term contributions to stop and sbottom masses,
\begin{eqnarray}
\label{approx1}
\Delta^2 & = &
\frac{3 g^2}{64 \pi^2 \sin^2 \beta \cos^2 \beta m_W^2}
\nonumber \\
& & \nonumber \\
& & \times
\left\{ \frac{(m_b^2-m_W^2 \cos^2 \beta)(m_t^2-m_W^2 \sin^2 \beta)}
{\msta^2-\msba^2} \left[ f(\msta^2) - f(\msba^2) \right]
\right. \nonumber \\
& & + \left. \frac{m^2_t m^2_b}{\mstb^2 - \msbb^2}
\left[ f(\mstb^2) - f(\msbb^2) \right]
- \frac{2 m^2_t m^2_b}{m_t^2 - m_b^2}
\left[ f(m_t^2) - f(m_b^2) \right] \right\}
\end{eqnarray}
and
\begin{equation}
\label{effe}
f (m^2) = 2 m^2 \left( \log \frac {m^2}{Q^2} - 1 \right).
\end{equation}
The most striking fact in eqs. (\ref{cpeven1})--(\ref{effe}) is that the
correction $\Delta_2^2$ is proportional to $(m_t^4/m_W^2)$. This implies
that, for $m_t$ in the range of eq. (\ref{mtop}), the tree-level predictions
for $m_h$ and $m_H$ can be badly violated, and so for the related
inequalities. The other free parameter is $\msq$, but the dependence
on it is much milder. To illustrate the impact of these results, we display
in fig.~1 contours of the maximum allowed value of $m_h$ (reached for $m_A
\to \infty$), in the $(m_t,\tb)$ and $(m_t,\msq)$ planes, fixing $\msq =
1 \tev$ and $\tb=m_t/m_b$, respectively. In the following, when making
numerical examples we shall always choose the representative value $\msq
= 1 \tev$. To plot different quantities of physical interest
in the $(m_A,\tb)$ plane, which is going to be the stage of the following
phenomenological discussion, one needs to fix also the value of $m_t$.
In this paper, whenever an illustration of the $m_t$ dependence is needed,
we work with the two representative values $m_t=120,160 \gev$, which are
significantly different but well within the range of eq. (\ref{mtop}).
Otherwise, we work with the single representative value $m_t=140 \gev$.
As an example, we show in figs.~2--4 contours of constant $m_h$, $m_H$, and
$m_{H^\pm}$ in the $(m_A,\tb)$ plane. Here and in the following we vary
$m_A$ and $\tan \beta$ in the ranges
\be
\label{bounds}
0 \le m_A \le 500 \gev,
\;\;\;\;
1 \le \tan \beta \le 50 \, .
\ee

The effective-potential method allows us to compute also the leading
corrections
to the trilinear and quadrilinear Higgs self-couplings. A detailed discussion
and the full diagrammatic calculation will be given elsewhere. Here we
just give the form of the leading radiative corrections to the
trilinear $hAA$, $HAA$, and $Hhh$ couplings, which will play an important
role in the subsequent discussion of Higgs-boson branching ratios.
One finds [\ref{berz},\ref{brignole2}]
\begin{equation}
\label{hhh1}
\lambda_{hAA}
=
\lambda_{hAA}^0
+
\Delta \lambda_{hAA} \, ,
\;\;\;\;
\lambda_{HAA}
=
\lambda_{HAA}^0
+
\Delta \lambda_{HAA} \, ,
\;\;\;\;
\lambda_{Hhh}
=
\lambda_{Hhh}^0
+
\Delta \lambda_{Hhh} \, ,
\end{equation}
where
\begin{equation}
\label{lhaa0}
\lambda_{hAA}^0
=
- {i g m_Z \over {2 \cos \theta_W}}
\cos 2 \beta \sin (\beta + \alpha) \, ,
\end{equation}
\be
\label{lhhaa0}
\lambda_{HAA}^0
=
\phantom{-} {i g m_Z \over {2 \cos \theta_W}}
\cos 2 \beta \cos (\beta + \alpha) \, ,
\end{equation}
\be
\label{lhhh0}
\lambda_{Hhh}^0
=
- {i g m_Z \over {2 \cos \theta_W}}
[ 2 \sin (\beta + \alpha) \sin 2 \alpha
- \cos (\beta + \alpha) \cos 2 \alpha] \, ,
\end{equation}
and, neglecting the bottom Yukawa coupling and the $D$-term contributions
to squark masses
\begin{equation}
\label{deltahaa}
\Delta \lambda_{hAA}
=
- {i g m_Z \over {2 \cos \theta_W}}
  {3 g^2 \cos^2 \theta_W \over {8 \pi^2}}
  {\cos  \alpha \cos^2 \beta \over  {\sin^3 \beta}}
  {m_t^4 \over m_W^4}
  \log
  {\msq^2 + m_t^2 \over m_t^2},
\end{equation}
\begin{equation}
\label{deltahhaa}
\Delta \lambda_{HAA}
=
- {i g m_Z \over {2 \cos \theta_W}}
  {3 g^2 \cos^2 \theta_W \over {8 \pi^2}}
  {\sin  \alpha \cos^2 \beta \over  {\sin^3 \beta}}
  {m_t^4 \over m_W^4}
  \log
  {\msq^2 + m_t^2 \over m_t^2},
\end{equation}
\begin{equation}
\label{deltahhh}
\Delta \lambda_{Hhh}
=
- {i g m_Z \over {2 \cos \theta_W}}
  {3 g^2 \cos^2 \theta_W \over {8 \pi^2}}
  {\cos^2  \alpha \sin \alpha \over  {\sin^3 \beta}}
  {m_t^4 \over m_W^4}
  \left( 3 \log   {\msq^2 + m_t^2 \over m_t^2}
       - 2 {\msq^2 \over \msq^2 + m_t^2} \right).
\end{equation}
Notice that, besides the obvious explicit dependence, in
eqs.~(\ref{hhh1})--(\ref{deltahhh}) there is also an important implicit
dependence on $m_t$ and $\msq$, via the angle $\alpha$, which is determined
from the mass matrix of eqs.~(\ref{cpeven1})--(\ref{delta2}).
We also emphasize that neglecting the $D$-terms in the stop and sbottom
mass matrices is guaranteed to give accurate results only for $m_t
\gg m_Z$. For $m_t \sim m_Z$, one should make sure that the inclusion
of $D$-terms does not produce significant modifications of our results.
In the case of the $h$ and $H$ masses, and of the mixing angle $\alpha$,
complete formulae are available, and this check can be easily performed.
In the case of the $hAA$, $HAA$ and $Hhh$ couplings, complete formulae are
not yet available. For the phenomenologically most important coupling
at the LHC and SSC, $\lambda_{Hhh}$, we have explicitly checked that
the inclusion of $D$-terms does not produce important modifications
of our results.

Finally, one should consider Higgs couplings to vector bosons and fermions.
Tree-level couplings to vector bosons are expressed in terms of gauge
couplings and of the angles $\beta$ and $\alpha$. The most important part
of the radiative corrections is taken into account by using
one-loop-corrected formulae to determine $\alpha$ from the input parameters.
Other corrections are at most of order $g^2 m_t^2 / m_W^2$ and can be
safely neglected for our purposes. Tree-level couplings to fermions are
expressed in terms of the fermion masses and of the angles $\beta$ and
$\alpha$. In this case, the leading radiative corrections can be taken
into account by using the one-loop-corrected expression for $\alpha$ and
running fermion masses, evaluated at the scale $Q$ which characterizes
the process under consideration. This brings us to the discussion of the
renormalization group evolution of the top and bottom Yukawa couplings
in the MSSM. As boundary conditions, we assume as usual that $m_t(m_t)
=m_t$ and $m_b(m_b)=m_b$, with $m_b=4.8 \gev$ and $m_t$ numerical
input parameters. As stated in the Introduction, we assume in this
paper that all supersymmetric particles are heavy.
Then, since we want to compute Higgs-boson production cross-sections and
branching ratios, we are interested in the standard renormalization group
evolution of $h_t(Q)$ [$h_b(Q)$] from $Q=m_t$ [$Q=m_b$] to $Q \simeq
m_{H^\pm},m_H$, which is dominated by gluon loops.

To illustrate the behaviour of the Higgs couplings to vector bosons and
fermions, as functions of the input parameters, we show in figs.~5--7
contours in the $(m_A,\tb)$ plane of some of the correction factors
appearing in table~\ref{couplings}.
\section{LEP limits and implications}
In this section, we briefly summarize the implications of the previous results
on MSSM Higgs boson searches at LEP~I and LEP~II.
Partial results were already
presented in refs. [\ref{bf},\ref{erz3}].

As already clear from tree-level analyses, the relevant processes
for MSSM Higgs boson searches at LEP I are $Z \to h Z^*$ and $Z \to h A$,
which play a complementary role, since their rates are proportional to
$\sabsq$ and $\cabsq$, respectively. An important effect of radiative
corrections [\ref{berz}] is to allow, for some values of the
parameters, the decay $h \to AA$, which would be kinematically forbidden
according to tree-level formulae. Experimental limits which take radiative
corrections into account have by now been obtained by the four LEP
collaborations [\ref{lep}], using different methods to present and
analyse the data, and different ranges of parameters in the evaluation of
radiative corrections. A schematic representation of the presently
excluded region of the $(m_A,\tb)$ plane, for the standard parameter
choices discussed in sect.~2, is given in fig.~8, where the solid lines
correspond to our na\"{\i}ve\footnote{
We fitted the experimental exclusion contours, corresponding to $m_t = 140
\gev$ and the other parameters as chosen here, with two numerical values
for $\Gamma(Z \to h Z^*)$ and $\Gamma(Z \to h A)$. We have then computed
radiative corrections for the two values of $m_t$ considered here,
assuming that the variations in experimental efficiencies are small
enough not to affect our results significantly.} extrapolation of the
exclusion contour given in the first of refs. [\ref{lep}]. For a discussion of
the precise experimental bounds, we refer the reader to the above-mentioned
experimental publications.

The situation in which the impact of radiative corrections is most
dramatic is the search for MSSM Higgs bosons at LEP II. At the time
when only tree-level formulae were available, there was hope that LEP
could completely test the MSSM Higgs sector. According to tree-level formulae,
in fact, there should always be a CP-even Higgs boson with mass smaller than
($h$) or very close to ($H$) $m_Z$, and significantly coupled to the $Z$
boson.
However, as should be clear from the previous section, this result can be
completely upset by radiative corrections. A detailed evaluation of the
LEP II discovery potential can be made only if crucial theoretical parameters,
such as the top-quark mass and the various soft supersymmetry-breaking masses,
and experimental parameters, such as the centre-of-mass energy, the luminosity
and the $b$-tagging efficiency, are specified. Taking for example $\sqrt{s} =
190 \gev$, $m_t \simgt 110 \gev$, and our standard values for the soft
supersymmetry-breaking parameters, in the region of $\tan \beta$ significantly
greater than 1, the associated production of a $Z$ and a CP-even Higgs can be
pushed beyond the kinematical limit. Associated $hA$ production could be a
useful complementary signal, but obviously only for $m_h+m_A< \sqrt{s}$.
Associated $HA$ production is typically negligible at these energies.
To give a measure of the LEP II sensitivity, we plot in fig.~8 contours
associated with two benchmark values of the total cross-section $\sigma(
e^+ e^- \to hZ^*, HZ^*, hA, HA)$. The dashed lines correspond to $\sigma =
0.2 \, {\rm pb}$ at $\sqrt{s} = 175 \gev$, which could be seen as a rather
conservative estimate of the LEP II sensitivity. The dash-dotted lines
correspond to $\sigma = 0.05 \, {\rm pb}$ at $\sqrt{s} = 190 \gev$, which
could be seen as a rather optimistic estimate of the LEP II sensitivity.
In computing these cross-sections, we have taken into account the finite
$Z$ width, but we have neglected initial state radiation, which leads to
suppression near threshold. A more accurate estimate of the LEP II
sensitivity can be found in ref. [\ref{janot}].
Of course, one should keep in mind that there is, at least in principle,
the possibility of further extending the maximum LEP energy up to values
as high as $\sqrt{s} \simeq$ 230--240 GeV, at the price of introducing more
(and more performing) superconducting cavities into the LEP tunnel
[\ref{treille}].

In summary, a significant region of the parameter space for MSSM Higgses
could be beyond the reach of LEP II, at least if one sticks to the reference
centre-of-mass energy $\sqrt{s} \simlt 190 \gev$. The precise knowledge of
this region is certainly important for assessing the combined discovery
potential of LEP and LHC/SSC, but it does not affect the motivations and
the techniques of our study, devoted to LHC and SSC searches. Whether
or not a Higgs boson will be found at LEP in the future, we want to
investigate the possibilities of searching for all the Higgs states of the
MSSM at large hadron colliders, in the whole region of parameter
space which is not already excluded at present. Even if a neutral Higgs
boson is found at LEP, with properties compatible with the SM Higgs boson
within the experimental errors, it will be impossible to exclude that it
belongs to the MSSM sector. The LHC and SSC could then play a role in
investigating its properties and in looking for the remaining states of
the MSSM.

Similar considerations can be made for charged-Higgs searches at LEP~II
with $\sqrt{s} \simlt 190 \; {\rm GeV}$. In view of the $\beta^3$ threshold
factor in $\sigma ( e^+ e^- \to H^+ H^-)$, and of the large background from
$e^+ e^- \to W^+ W^-$, it will be difficult to find the $H^{\pm}$ at LEP II
unless $\mch \simlt m_W$, and certainly impossible unless $\mch < \sqrt{s}/2$.
We also know [\ref{berz},\ref{higgscorr}] that for
generic values of the parameters there are no large negative radiative
corrections to the charged-Higgs mass formula, eq. (\ref{mch0}). A comparison
of figs.~4 and 8 indicates that there is very little hope of finding the
charged Higgs boson of the MSSM at LEP II (or, stated differently, the
discovery of a charged Higgs boson at LEP II would most probably rule out
the MSSM).
\section{Branching ratios}
\subsection{Neutral Higgs bosons}
The branching ratios of the neutral Higgs bosons of the MSSM were
systematically studied in ref. [\ref{kz}], using the tree-level
formulae for masses and couplings available at that
time\footnote{Also, the partial widths for the decays $h,H,A
\rightarrow\gamma\gamma$ were affected by numerical errors.}
(previous work on the subject is summarized in ref. [\ref{hunter}]).
Here we present a systematic study which includes the radiative corrections
described in sect.~2. As usually done for the SM Higgs boson, we
consider the two-body decay channels
\be
\label{smdecays}
h,H,A \longrightarrow
c\overline{c}, \
b\overline{b}, \
t\overline{t}, \taptam, gg, \gamma\gamma,\
W^*W^*,\  Z^*Z^*,\  Z\gamma \, .
\ee
For consistency, we must also consider decays with one or two Higgs bosons
in the final state
\be
\label{newdecays}
h \rightarrow AA,
\ \
H \rightarrow hh, \, AA, \, ZA,
\ \
A\rightarrow Zh.
\ee
On the other hand, we neglect here possible decays of MSSM Higgs bosons
into supersymmetric particles: as previously stated, we consistently assume
a heavy spectrum of $R$-odd particles, so that only $R$-even ones can be
kinematically accessible in the decays of $h,H,$ and $A$. We perform our
study in the framework of MSSM parameter space, with the representative
parameter choices illustrated in sect.~2. The effects of changing the mass
of the top quark, and the sensitivity to squark masses in the high-mass
region, will also be briefly discussed.

The partial widths for the decays of eq. (\ref{smdecays}) that correspond
to tree-level diagrams can be obtained from the corresponding formulae
for the SM Higgs boson (for a summary, see ref.~[\ref{hunter}]), by simply
multiplying the various amplitudes by the supersymmetric correction
factors listed in table~1.
For decays that are described by loop diagrams, however, in the MSSM
one has to include some contributions that are absent in the SM.
Diagrams corresponding to the exchange of $R$-odd supersymmetric particles
give negligible contributions to the
corresponding partial widths, in the limit of heavy supersymmetric-particle
masses that we have chosen for our analysis (in accordance with intuitive
ideas about decoupling). One must also include
the charged-Higgs loop contributions to the $\gamma \gamma$ and $Z \gamma$
final states. When considering instead the processes of eq. (\ref{newdecays}),
we improve the tree-level formulae of ref. [\ref{hunter}]
not only with the self-energy corrections to the mixing angle $\alpha$,
but also with the vertex corrections of eqs. (\ref{hhh1})--(\ref{deltahhh}).

QCD [\ref{gorishnyh}] and electroweak [\ref{kniehlbardin}]  ra\-diative
cor\-rections  to the fermion-anti\-fermion and the $WW$,\ $ZZ$ channels
have been recently computed for the SM Higgs boson, $\varphi$. They have
been found to be small (less than $\sim 20\%$), with the exception of
the QCD corrections to the decays into charm and bottom quark pairs, which
are large because of running-quark-mass effects.
We then included the QCD corrections as described in ref. [\ref{zkwjsaa}].
One may also wonder
whether running-mass effects induced by the large top Yukawa coupling
could give further important effects. However, one can easily see that
these effects give corrections which are certainly less than 20\%.

The QCD correction to $\varphi \to \gamma\gamma$ is also available, and
known to be negligibly small [\ref{djouadigamgam}].
Sizeable QCD corrections are found, however, for the decay $\varphi
\to gg$ [\ref{djouadigg}].
Although this effect is not important for the branching ratio study, since
$\varphi \to gg$ is neither the dominant decay mode nor a useful channel for
detection, it still has to be included in the production cross-section of $h$
via the two-gluon fusion mechanism.

Another general and well-known property of the MSSM is that the
self-interactions of the Higgs bosons are controlled, modulo the
logarithmic corrections discussed in sect.~2, by the $SU(2)$ and
$U(1)$ gauge couplings. Therefore, the total widths of {\it all}
MSSM Higgs bosons, displayed in fig.~9,
stay below 10 GeV in the whole parameter space we have considered.

The most important branching ratios for the neutral MSSM Higgs bosons are
shown, as a function of the mass of the decaying particle, in figs.~10--12.
To avoid excessive proliferation of figures, we consider the two
representative values
\be
\label{tbvalues}
\tb= 1.5, \, 30,
\ee
and for each of these we vary $m_A$ between the experimental lower bound
of fig.~8 ($m_A \simeq 59 \gev$ for $\tb=1.5$, $m_A \simeq 44 \gev$ for
$\tb=30$) and the upper bound of eq. (\ref{bounds}), assuming $m_t=140\gev$
and  $m_{\tilde q}=1\tev$.

We consider first the branching ratios of $h$ (fig.~10). We can clearly see
the effect of radiative corrections on the allowed range of $m_h$ for the
given values of $\tb$. For $m_A \simlt 25 \gev$, the decay $h \to AA$ can
be kinematically allowed and even become the dominant mode. This decay channel
was important at LEP I, but since the corresponding region of parameter space
is already excluded by experiment, this decay mode does not appear in fig.~10.
The dominant decay mode is then $h\to b\bar{b}$, whereas the $\taptam$ mode
has a branching ratio of about $8\%$ throughout the relevant part of the
parameter space. In fig.~10, one immediately notices the rather steep slopes
for the $c\bar{c}$ and $\gamma\gamma$ branching ratios plotted versus $m_h$,
with larger effects for larger values of $\tb$: their origin can be understood
by looking at figs.~2 and 5--7, which show how $m_h$ and the $h$ couplings
to heavy fermions and vector bosons vary in the $(m_A,\tb)$ plane.

If the SM Higgs boson is in the intermediate mass region, $m_{\varphi} =$
70--140 GeV, at large hadron colliders a measurable signal can be obtained
via the $\gamma\gamma$  mode. Since the mass of the light Higgs $h$ is
indeed below or inside this region, the $\gamma\gamma$ mode is also crucial
for the MSSM Higgs search. Furthermore, the $\gamma \gamma$ branching ratio
as a function of the Higgs mass exhibits a rather peculiar behaviour, not only
for $h$ but also for $H$ and $A$, so a more detailed discussion is in order.
The partial width is given by
\be
\label{gammaw}
\Gamma(\phi\to\gamma\gamma)
={\alpha^2 g^2\over 1024\pi^3}{m_{\phi}^3\over m_W^2}
\abs{\sum_i I_i^{\phi}(\tau^\phi_i)}^2 , \qquad
\tau^{\phi}_i={4m_i^2\over m_{\phi}^2},
\ee
where $\phi=h,H,A$ and $i=f,W,H,\tilde{f},\tilde{\chi}$ indicates
the contributions from ordinary fermions, charged gauge bosons, char\-ged
Higgs bosons, sfermions and charginos, respectively. The functions
$I_\phi^i(\tau^i_{\phi})$ are given by
\bea
\label{ifunct}
I_f^{\phi} & = & F_{1/2}^{\phi}(\tau^{\phi}_f)N_{cf}e^2_f\ R_f^{\phi} \, ,
\nonumber \\
I_W^{\phi} & = & F_{1}(\tau^{\phi}_W) R_W^{\phi} \, ,
\nonumber \\
I_H^{\phi} & = & F_{0}(\tau^{\phi}_H) R_H^{\phi} {m_W^2\over m_{\phi}^2} \, ,
\nonumber \\
I_{\tilde{f}}^{\phi} & = & F_{0}(\tau^{\phi}_{\tilde{f}})
N_{cf}e^2_f  R_{\tilde{f}}^{\phi}{m_Z^2\over m_{\tilde{f}}^2},
\nonumber \\
I_{\tilde{\chi}}^{\phi} & = & F_{1/2}^{\phi}(\tau^{\phi}_{\chi})
R_{\tilde{\chi}}^{\phi}{m_W\over m_{\tilde{\chi}}},
\eea
where $N_{cf}$ is $1$ for (s)leptons and $3$ for (s)quarks, and the
subscripts of the complex functions $F_{1/2}^S(\tau)$,
$F_{1/2}^P(\tau)$, $F_{0}(\tau)$, and $F_1(\tau)$, which were
calculated in ref. [\ref{vvzs}], indicate the spin of the
particles running in the loop. In the case of spin-$1/2$
particles, the contribution is different for CP-even and CP-odd
neutral Higgses. The symbols $R^{\phi}_i$ denote the appropriate
correction factors for the MSSM Higgs couplings: for $i=f,W$ they are
given in table 1, for $i=H,\tilde{f},\tilde{\chi}$ they can be found,
for example, in Appendix C of ref. [\ref{hunter}]. The $W$ contribution
dominates the $h \to \gamma \gamma$ decay rate. The function $F_1$ is
large at and above $\tau=1$. For the $W$ contribution $\tau =4m_W^2/m_h^2
> 1$, and increasing $m_h$ gives increasing values of $F_1$. The steep
dependence of the branching ratio on $m_h$ is a consequence of the
fast change of $\sabsq$ as $m_A$ is increased for fixed $\tb$. This is
further enhanced by the fact that the large interval $100 \gev \simlt
m_A \le 500\gev$ is mapped into a very small interval (a few GeV)
in $m_h$. We elucidate this effect by plotting in fig.~13 the
branching ratios  of $h$ as a function of $m_A$, for the same values of
the parameters as in fig.~10. We can see that the tip of the $gg, c \ov{c}$
and $\gamma \gamma$ curves in fig.~10 is mapped into a long
plateau in fig.~13. We can also observe that in a large
region of the parameter space the $h \to \gamma \gamma$ branching ratio
has a value somewhat smaller than (but comparable to) the corresponding
branching ratio for a SM Higgs of mass $m_h$. This is due to the
fact that  all the $h$ couplings tend to the SM Higgs couplings for
$m_A \gg m_Z$; however, for the $h$ couplings to fermions the approach to
the asymptotic value is much slower than for the $h$ couplings to
vector bosons, as can be seen from figs.~5 to 7.
In fig.~13, the branching ratios for the $W^*W^*$
and $Z^*Z^*$ decays are also plotted, whereas they were omitted in
fig.~10 in order to avoid excessive crowding of curves. However,
for our parameter choice they have little interest at large hadron
colliders, because of the small production rates and the large backgrounds.

The branching ratios of the heavy Higgs boson $H$, depicted in fig.~11,
have a rather complicated structure. We make here four remarks.

i)
The $\gamma\gamma$
mode has a steeply decreasing branching ratio with increasing $m_H$,
except at small values of $\tb$ and at the lower kinematical limit of $m_H$,
where one or more of the $AA$, $ZA$ and $hh$ decay channels are open.
The steep fall of the $\gamma\gamma$ branching ratio at large values of $\tb$
can be easily understood. The partial width $\Gamma (H \to\gamma\gamma)$
is dominated by the $W$ contribution, proportional to $\cos^2(\beta-\alpha)$.
As we can see in fig.~7, $\cos^2(\beta-\alpha)$ decreases very fast, for
increasing $m_H$, at fixed values of $\tb$. This steep decrease is slightly
compensated by the increase of $F_1(\tau_W)$ at $\tau_W \leq 1$,
which has a peak at the $W$ threshold $m_H=2 m_W$. Another peak in the
$\gamma \gamma$ branching ratio is obtained, for small values of $\tb$,
at $m_H = 2 m_t$, where the top-quark loop gives the dominant contribution.

ii)
The complicated structure in the $H$ branching ratio curves is mainly due to
the $H\to hh$ channel. For $m_H<2 m_t$, and not too high values of $\tb$, this
decay mode is dominant whenever kinematics allows.  This channel is always
open at the lower kinematical limit of $M_H$. Increasing $M_H$ a little bit,
however, it may become strongly suppressed, because for small
increasing values of $m_A$ the value of $m_h$ rises faster than that of $m_H$,
so that the channel can become kinematically closed. Obviously, for
sufficiently high values of $m_H$ the channel is always open.
At high values of $\tb$, the mass region at the lower kinematical limit
where $H\to hh$ is open becomes smaller and smaller, explaining the presence
of the almost vertical line in fig.~11.
A further structure is present in this decay channel due to the coupling
factor $\lambda_{Hhh}$ [see eqs. (\ref{lhhh0}) and (\ref{deltahhh})]\@.
There are relatively small values of $m_H$ at which $\lambda_{Hhh}$
accidentally vanishes. Furthermore, for very large values of $m_H$ and $\tb$
one has $\alpha \simeq 0$, $\beta \simeq \pi/2$, and therefore $\lambda_{Hhh}
\simeq 0$. Unfortunately, even when it is dominant, this mode has very large
backgrounds, so it seems unlikely to give a measurable signal at large
hadron colliders. The $H \to AA$ mode is kinematically allowed only for
values of $m_A$ below 50--60 GeV, in which case it can have a large branching
ratio, competing with the one for $H \to hh$. The $H \to Z A$ mode is
kinematically allowed only in the region of parameter space which is already
excluded by the LEP I data.

iii)
$H$ can decay at tree level into $ZZ\to l^+l^-l^+l^-$, which is the
`gold-plated' signature for the SM Higgs boson. Unfortunately, in the case
of $H$ the branching ratio is smaller, and it decreases fast with increasing
$\tb$ and/or $m_A$. For small $\tb$ and $2 m_h < m_H < 2 m_t$, this mode is
suppressed by the competition with $H \to hh$, and this effect is further
enhanced by the inclusion of the radiative correction of eq. (\ref{deltahhh}),
which typically gives an additional $50 \%$ suppression.
Nevertheless, as we shall see in the next section, the four-lepton channel
can give a measurable signal in some small region of the parameter space.

iv) The decay into $t\bar{t}$ is dominant above threshold at moderate values
of $\tb$. But above $\tb\sim 8$ or so the $b\bar{b}$ mode remains dominant
and $\taptam$ has the typical $\sim 10\%$ branching ratio.

Finally, we discuss the branching ratios of $A$, shown in fig.~12.
The $\gamma\gamma$ one is always small, although at small $\tb$
and slightly below the top threshold, $m_A\sim 2 m_t$, it reaches a value
$\sim 8 \times 10^{-4}$, which may give a measurable signal in a small
island of the parameter space. The behaviour of the $A \to \gamma\gamma$
branching ratio can be easily understood by taking into account that the
partial width is dominated by the top loop contribution.
Two features are important here. First, the function $F_{1/2}(\tau)$
appearing in eq. (\ref{ifunct}) has a strong enhancement at $\tau\sim 1$.
Furthermore, the $t\bar{t}A$ coupling gives a suppression factor $1/(\tb)^{2}$
for increasing values of $\tb$.

At smaller values of $m_A$ and $\tb$, there is a substantial branching
ratio to $Zh$, which however does not look particularly promising for
detection at large hadron colliders, because of the very large $Zb\bar{b}$
background. We can see that all the dominant decay modes of the $A$ boson
correspond to channels which are overwhelmed by very large background,
except perhaps the $\taptam$ mode, which, as we shall see in the next
section, may give a detectable signal for very high values of $\tb$.

We have studied the neutral Higgs branching ratios also at $m_t=120,160,180
\gev$. Increasing the top mass has two major effects. First, the maximum
value of $m_h$ increases (see fig.~1). Next, owing to the increased value of
the top threshold, the structure generated by the opening of the top decay
channel is shifted to higher mass values.
We also note that varying $m_{\tilde{q}}$ in the range 0.5--2 TeV has
negligible effects on the branching ratio curves of figs.~10--12.
Finally, if one chooses $m_t$ and $\msq$ so large that $m_h > 130
\gev$, the $W^*W^*$ and $Z^*Z^*$ branching ratios can become relevant
also for $h$.

\subsection{Charged Higgs boson}

In the case of the charged Higgs boson, we considered only the two-body
decay channels
\be
\label{chchan}
H^+ \to c \ov{s}\, , \; \tau^+ \nu_{\tau} \, , \; t \ov{b} \, , \;
W^+ h \, .
\ee
Tree-level formulae for the corresponding decay rates can be found, for
instance,
in ref. [\ref{hunter}]. Loop-induced decays such as $H^+ \to W^+ \gamma,
W^+ Z$ have very small branching ratios [\ref{loopch}] and are not relevant
for experimental searches at the LHC and SSC. Radiative corrections to
the charged-Higgs-boson mass formula were included according to eqs.
(\ref{sumrule}) and (\ref{approx1}). The $H^+ W^- h$ coupling, proportional
to $\cos (\beta - \alpha)$, was evaluated with the one-loop corrected value
of $\alpha$. The leading QCD corrections to the $H^+ b \ov{t}$ and
$H^+ s \ov{c}$ vertices  were parametrized, following refs. [\ref{qcdch}],
by running quark masses evaluated at a scale $Q \sim \mch$. The resulting
branching ratios for the charged Higgs boson are displayed in fig.~14, for
$\tb = 1.5, \, 30$ and the standard parameter choice $m_t = 140 \gev$, $\msq
= 1 \tev$. One can see that the dominant factor affecting the branching ratios
is the $\mch = m_t+m_b$ threshold.
Above threshold, the $t \ov{b}$ mode is dominant
for any value of $\tb$ within the bounds of eq. (\ref{bounds}).  Below
threshold, the dominant mode is $\tau^+ \nu_{\tau}$, with the competing mode
$c \ov{s}$ becoming more suppressed for higher values of $\tb$. For small
values of $\tb$, the $W^+ h$ decay mode can also be important, and even
dominate, in a limited $\mch$ interval, if the $W^+ h$ threshold opens up
before the $t \ov{b}$ one. The exact position of the two thresholds on
the $\mch$ axis depends of course on $\tb$, $m_t$, and $\msq$. It is just a
numerical coincidence that in fig.~14a the two thresholds correspond
almost exactly.
For increasing values of $\mch$ and $\tb$, the numerical relevance of the
$W^+ h$ branching ratio rapidly disappears, because of the $\cos^2 (\beta
- \alpha)$ suppression factor in the corresponding partial width.

The total  charged Higgs boson width is shown, as a function of $\mch$ and for
$\tb = 1.5, \, 3, \, 10, \, 30$, in fig.~9d. Again one can see the effects of
the $t \ov{b}$ threshold, and also the $\tb$-dependence of the couplings to
fermions. In any case, the charged Higgs width remains smaller than 1 GeV for
$\mch < m_t + m_b$, and smaller than $10 \gev$ for $\mch < 500 \gev$.
\section{Neutral-Higgs production cross-sections}

There is only a limited number of parton-level processes which can give
interesting rates for the production of the MSSM neutral Higgs bosons
($\phi=h,H,A$) at proton-proton supercolliders:
\bea
\label{prodmech}
g+g \to \phi \, ,\\
q + q \to q + q+  W^* + W^* \to q + q + \phi \, , \\
g+g \ \ {\rm or\ \ } q + \overline{q}\to b+\overline{b}+\phi \, ,\\
g+g \ \ {\rm or\ \  } q + \overline{q}\to t+\overline{t}+\phi \, , \\
q+ \ov{q} \to W(Z) + \phi \, ,
\eea
where $q$ denotes any quark flavour.
These processes are controlled by the Higgs couplings to heavy quarks and
gauge bosons, whose essential features were summarized in table~1. We
briefly discuss here the corresponding cross-sections and the status of
their theoretical description, emphasizing the features which are
different from the SM case. We shall always adopt the HMRSB structure
functions [\ref{hmrs}] with $\Lambda^{(4)} = 190 \mev$.
\vskip 0.5cm
{\bf Gluon fusion. \ }
In the SM, $g g \to \phi$ [\ref{georgi}] is the dominant production
mechanism, the most
important diagram being the one associated with the top-quark loop. In the
MSSM, this is not always the case, since the correction factors of table~1
give in general suppression for the top contribution and enhancement for
the bottom one, and stop and sbottom loops could also play a role.

The leading-order amplitudes for the gluon-fusion processes
are determined by the functions of eqs. (\ref{ifunct}), with top,
bottom, stop and sbottom intermediate states. For $\msq=1\tev$, the squark
contributions are very small, owing to the suppression factor $m_Z^2
/ \msq^2$ in the corresponding $I_{\tilde{q}}$ functions. For large values
of $\tb$, the bottom contribution can compete with the top one and even
become dominant.

QCD corrections to the gluon-fusion cross-section were recently evaluated
in ref. [\ref{djouadigg}], for a SM Higgs in the mass region below the
heavy-quark threshold. In this region, QCD corrections increase the top
contribution by about $50\%$. To a good approximation, the bulk of QCD
corrections can be taken into account by performing the replacement
\be
\label{qcdgg}
\sigma_0 (g g \to \phi)
\longrightarrow
\sigma_0 (g g \to \phi)
\left[
1 + \left( \frac{11}{2} + \pi^2 \right) \frac{\alpha_S}{\pi}
\right],
\ee
at the renormalization scale $Q = m_{\phi}$. This calculation, unfortunately,
is not valid above the heavy-quark threshold, a region which is relevant for
the bottom contribution and for the top contribution to $H,A$ production, when
$m_H,m_A>2 m_t$. Even below the heavy-quark threshold, the SM QCD corrections
are applicable to $h$ and $H$ production, but not to $A$ production, because
of the additional $\gamma_5$ factor appearing at the $A q \ov{q}$ vertex.
In view of this not completely satisfactory status of QCD corrections, we
calculate, conservatively, the top contribution without QCD corrections.
However, when discussing the detectability of the different physics signals,
we shall take into account the results of ref. [\ref{djouadigg}], when
applicable. In the case of the bottom contribution, we use the running
$m_b$, which leads to suppression.

In fig.~15 we display cross-sections for $gg \to \phi$ ($\phi=h,H,A$),
as functions of $m_{\phi}$, for $\tb=1.5,3,10,30$ and for LHC and SSC
energies. The SM Higgs cross-section is also shown for comparison. For
large values of $\tb$ and not too high values of $m_{\phi}$, the
cross-sections can be enhanced  with respect to the SM value.
This effect is due to the enhanced bottom-quark contribution,
as apparent from table~1 and fig.~5. The fast disappearance of
this effect for increasing Higgs masses is due to the fast decrease
of the function $F_{1/2}^{\phi} ( \tau_b^{\phi})$ as $\tau_b^{\phi}
\to 0$. When the neutral Higgs couplings to fermions are SM-like,
the gluon-fusion cross-sections approach the SM value, and are always
dominated by the top contribution. The changes in the slopes of the
curves in fig.~15 are due to the competing top and bottom contributions.
In particular, one can notice an important threshold effect, for $m_A
\sim 2 m_t$, in the process $g g \to A$, which can bring the corresponding
cross-section above the SM value for low $\tb$.

As a final remark, we notice that the LHC and SSC curves in fig.~15
have very similar shapes, with a scaling factor which is determined
by the gluon luminosity and uniformly increases from $\sim 2.5$ at
$m_{\phi} \sim 100 \gev$ to $\sim 5$ at $m_{\phi} \sim 500 \gev$.

\vskip 0.5cm
{\bf $W$ fusion.\ }
In the SM case, the $W$-fusion mechanism [\ref{wfusion}]
can compete with the gluon-fusion
mechanism only for a very heavy ($m_{\varphi} \simgt 500 \gev$) Higgs boson,
owing to the enhanced $W_LW_L\varphi$ coupling and to the relative increase of
the quark number-densities. In the MSSM, the correction factors for the
couplings to vector boson pairs (see table~1 and fig.~7) are always smaller
than 1, so that the MSSM $W$-fusion cross-sections are always smaller than
the SM one.

We illustrate this in fig.~16, where $W$-fusion cross-sections for $h$ and $H$
are displayed, for the same $\tb$  and $\sqrt{s}$ values as in fig.~15.
For both $h$ and $H$, the SM cross-section is approached from below in
the regions of parameter space where $\sin^2 (\beta - \alpha) \to 1$
and $\sin^2 (\beta - \alpha) \to 0$, respectively. In figs.~16b and 16d,
for $m_A \to 0$ there is a positive lower bound on $\sin^2 (\beta
- \alpha)$ (see fig.~7), reflecting the fact that at the tree level
$\alpha \to - \beta$ in this limit, so the SM value is actually never
reached. For increasing
$m_H$, one can notice the fast decoupling of $H$ from $W$-pairs, as already
observed when discussing the total width. In leading order, $A$ does not
couple to $W$-pairs. A non-vanishing cross-section could be generated at
one loop, but such a contribution is completely negligible, since even
for $h$ and $H$ the $W$-fusion cross-section is small ($\simlt 20 \%$)
compared with the gluon-fusion cross-section. Finally, we observe that
the LHC and SSC cross-sections of fig.~16 differ by an overall factor
$\sim 3$ in the phenomenologically relevant region, $m_{\phi} =$
70--140 GeV.

\vskip 0.5 cm
{\bf Associated $\bb \phi$ production. \ }
This mechanism is unimportant in the SM, since its  cross-section is
too small to give detectable signals [\ref{zktth}]. In the MSSM model,
however, for large values of $\tb$ the
$\bb \phi$ couplings can be strongly enhanced. Then for not too high
values of the Higgs masses, a significant fraction of the total cross-section
for neutral Higgs bosons can be due to this mechanism.

The associated $b \ov{b} \phi$ production involves two rather different mass
scales, $m_{\phi} >> m_b$, therefore at higher orders large logarithmic
corrections of order
$$
\alpha_s^n \ln^n \left( {m_{\phi}\over m_b} \right) \,
$$
may destroy the validity of the Born approximation, depending on the value
of $m_{\phi}$. One needs an
improved treatment where these logarithms are resummed to all orders.
The origin of these logarithms is well understood. Part of them
come from configurations where the gluons are radiated collinearly
by nearly on-shell bottom quarks, which are obtained by splitting
the initial gluons into a $\bb$ pair.
This type of logarithms are responsible for
the QCD evolution of the effective $b$-quark density within the proton:
they were carefully analysed and resummed to all orders,
and it was found [\ref{sophabgun}] that the corrections are positive
and increase with the Higgs-boson mass.
A second subset of logarithms lead to running quark mass effects.
An analysis where both effects are  treated simultaneously is still
missing. In view of this ambiguity, we interpolated the existing
results by using the Born approximation with the bottom quark mass
adjusted to the fixed value $m_b=4\gev$. However, one should keep in
mind that the theoretical estimate in this case has a large (factor of 2)
uncertainty.

In fig.~17 we display cross-sections for associated $\bb \phi$ production,
for the same $\tb$  and $\sqrt{s}$ values as in fig.~15.
Comparing the cross-sections of figs.~15 and 17, we can see that
the $h \bb$ cross-section can give at most a 20\% correction
to the total $h$ cross-section. The $H\bb$ and $A \bb$
cross-sections, however, can be even larger than the corresponding
gluon-fusion cross-sections for $\tb \simgt 10$.
Comparing the LHC and SSC curves of fig.~17, one can notice
a rescaling factor varying from $\sim 3$ to $\sim 8$ in the $m_{\phi}$
region from 60 to 500 GeV.

\vskip 0.5cm
{\bf Associated $W \phi$ ($Z \phi$) production. \ }
This mechanism [\ref{bjorken}] is the hadron collider analogue of the SM Higgs
production mechanism at LEP, with the difference that at hadron colliders
$W\phi$ production is more important than $Z\phi$ production. In the $Z\phi$
case,
the event rate at the LHC and SSC is too low to give a detectable signal,
both in the SM and (consequently) in the MSSM. The $W \phi$ mechanism has
considerable importance at the LHC for $\phi=h,H$ and in the Higgs mass
range $m_\phi=$ 70--140 GeV, where a measurable signal may be obtained
from final states consisting of two isolated photons and one isolated lepton.
The calculation of the cross-section is well understood, including
the QCD corrections, since it has a structure similar to the Drell-Yan
process, with the same next-to-leading-order corrections (for a recent
study concerning the numerical importance of the QCD corrections see
ref.~[\ref{willen}]). The QCD corrections are positive, and amount to
about 12\% if one chooses $Q^2=\hat{s}$ as the scale of $Q^2$ evolution.
The production cross-sections of $h$ and $H$ are obtained by rescaling
the SM model cross-section by the appropriate correction factors
given in table~1.

In fig.~18, cross-sections for $Wh$ and $WH$ are displayed,
as functions of corresponding Higgs masses,
for the same $\tb$  and $\sqrt{s}$ values as in fig.~15.
Since the SUSY correction factors are the same, the approach to the
SM case and the irrelevance of $WA$ production can be described in the same
way as for the $W$-fusion mechanism.

In the phenomenologically relevant region, $m_{\phi} =$ 70--140 GeV,
the scaling factor between the LHC and SSC curves is $\sim 2.5$.

\vskip 0.5cm
{\bf Associated $\tt \phi$ production. \ }
In the SM, the Born cross-section formula for this process is the same
as for the $b \ov{b} \varphi$ case [\ref{zktth}]. In the MSSM case,
one just needs to insert the appropriate SUSY correction factors, as
from table~1. Note, however, that the leading-order QCD calculation is
more reliable in this case, since in the $t \ov{t} \phi$ case
one does not have two very different physical scales when $m_{\phi}$
is in the intermediate mass region. The next-to-leading QCD corrections are
not known, therefore the Born cross-section still suffers from a relatively
large ($\sim 50\%$) scale ambiguity.

In fig.~19, the production cross-sections for $\tt \phi$ ($\phi=h,H,A$) are
plotted, as functions of the corresponding Higgs mass,
for the same $\tb$  and $\sqrt{s}$ values as in fig.~15.
In general, the MSSM cross-sections are
smaller than the SM one, which is approached in the limit
in which the $\tt \phi$ coupling becomes SM-like.
A possible exception is the $t \ov{t} H$ cross-section for small values
of $m_A$ and $\tb$, since in this case the corresponding coupling can
be slightly enhanced with respect to the SM one.

In the phenomenologically relevant range, $m_{\phi} =$ 70--140 GeV,
the rescaling factor between the LHC and SSC curves in fig.~19 varies
from $\sim 6$ to $\sim 7$.

\vskip 0.5 cm
In the phenomenologically allowed range of eq. (\ref{mtop}), the top-mass
dependence of the cross-sections of figs.~15--19 is not negligible, but it
does not change qualitatively the previous considerations. The largest
effect comes from the increase of the upper limit on $m_h$  for increasing
top mass (see fig.~1). This induces a shift in the limiting values for the
$h$ and $H$ production cross-sections. There are also obvious
kinematical top-mass
effects in the gluon-fusion mechanism and in the $\tt \phi$ mechanism, which
are well understood from SM studies [\ref{zkwjsaa}].
In the MSSM, additional effects are
given by the radiative corrections to the relevant Higgs couplings,
which were discussed in sect.~2.

\section{Physics signals}
\subsection{Neutral Higgs bosons}
We now calculate the rates for a number of processes that could
provide evidence for one or more of the neutral MSSM Higgs
bosons at the LHC and SSC, and we summarize our results with
the help of contour plots in the $(m_A,\tb)$ plane. We  consider
production cross-sections, folded with branching ratios, for the
following signals:
\begin{itemize}
\item
two isolated photons;
\item
one isolated lepton and two isolated photons;
\item
four isolated charged leptons;
\item
a pair of tau leptons.
\end{itemize}
In the SM case, the first two signals are relevant for the region of
intermediate Higgs mass, $70\gev < m_{\phi} < 140 \gev$, the third one is the
so-called `gold-plated' signal in the high-mass region $130\gev < m_{\phi}
< 800 \gev$, and the $\taptam$ signal appears to be hopelessly difficult.

In a complete phenomenological study, one would like to determine precisely
the statistical significance of the different physics signals. This would
require, besides the computation of total signal rates, the calculation of
the backgrounds, the determination of the
efficiencies (for both signals and backgrounds) due to kinematical cuts and
detector effects,  the optimization of the kinematical cuts to achieve the
best signal$/$background ratio, etc.
Such a complete analysis would require the specification of several detector
and machine parameters, and goes beyond the aim
of the present paper. Instead, we try here to present total rates for
well-defined physics signals, in a form which should be useful as a
starting point for dedicated experimental studies.

As the only exception, to illustrate with an example how our results can be
used to establish the statistical significance of a given physics signal in
a given detector, we shall describe the case of the `two-isolated-photons'
signal, using the results of recent simulation works.
A similar procedure should be adopted for
any other physics signal, detector, and collider, once complete results of
simulation works are available. In many cases, the existing results
from previous background and simulation studies, carried out for the SM,
can also be used to draw conclusions concerning the MSSM case. We mention,
however, two important differences: 1) the total widths of $H$ and $A$
remain small even in the high-mass region; 2) for large $\tb$, the number
of signal events in the $\taptam$ final state is significantly higher than
in the SM case.

\vskip 0.5cm
{\bf Inclusive two-photon channel.\ }
In fig.~20 we display cross-sections times branching ratios for the
inclusive production of $\phi=h,H,A$, followed by the decay $\phi \to
\gamma \gamma$, as functions of $m_{\phi}$, and for
the same parameter choices and energies as in fig.~15. We sum the
contributions of the gluon-fusion, $W$-fusion, and $\bb\phi$ mechanisms.
For comparison, the SM value is also indicated. The QCD corrections of
ref. [\ref{djouadigg}] are not included, for the reasons explained in
the previous section. In the case of $h$ and $H$,
the signal rates are always smaller than in the SM, and approach the SM
values at the upper and lower edge of the allowed $m_h$ amd $m_H$ ranges,
respectively. The rather steep slope characterizing the approach to the SM
limit, for varying Higgs mass and fixed $\tb$, is a reflection of the property
of the branching ratios discussed in sect.~4. Also the structure in figs.~20b
and 20e can be attributed to the threshold behaviour of the $H \to hh$
channel. The signal rate for the CP-odd $A$ boson is extremely small for
$\tb \simgt 3$. However, we observe that in a small
region of the parameter space, for $m_A$ just below $2 \, m_t$ and
$\tb \simlt 3$, the rate can become larger than the SM one: nevertheless,
in general it is still too low to produce a detectable signal, unless
one chooses $\tb \sim 1$ and $m_A \sim 2 m_t$.

For the inclusive two-photon channel, the results of detailed simulations
of signal and SM background are now available, for some of the LHC detector
concepts [\ref{cseez},\ref{evian},\ref{cseezcap},\ref{unal}]. For
illustrative purposes, in the following example we shall follow the
treatment of ref.~[\ref{cseezcap}]. In the mass range $m_\phi=$ 80--150 GeV,
and assuming $10^5 \, {\rm pb}^{-1}$ integrated luminosity, this LHC
simulation considers a fairly wide range of detector performances,
which affect the significance of the signal. For an energy resolution
$\Delta E / E = [2 \% / \sqrt{E(\!\!\gev)}] +0.5 \%$, ref.~[\ref{cseezcap}]
obtains a $10^4$ efficiency for rejecting jets faking an isolated photon in
the relevant $p_T$ region. Applying standard kinematical cuts, this simulation
finds $\sim$ 40--50$\%$ kinematical acceptance, with an additional $\sim$
30--40$\%$ loss due to isolation cuts and reconstruction efficiency for
the isolated photons. Typically, for a SM Higgs with $m_{\varphi} \sim
100 \gev$, one obtains $\sim 10^3$ signal events over $\sim 10^4$ background
events, corresponding to a statistical significance $S/\sqrt{B} \sim 10$.
More generally, ref. [\ref{cseezcap}] determined the statistical
significance of the signal for given values of the generic Higgs mass
$m_{\phi}$ and of the signal rate $\sigma \cdot BR (\phi \to \gamma \gamma)$
(see fig.~21). In our opinion, this is an excellent way of summarizing the
simulation work, since it gives the possibility of studying alternatives
to the SM case, and in particular the MSSM. The dashed line in fig.~21
corresponds to the signal for the SM Higgs, which includes both the
gluon-fusion and the $W$-fusion production mechanisms, and also
the QCD corrections of ref. [\ref{djouadigg}]. One can see from fig.~21
that for such optimistic detector parameters there is some margin for
detecting smaller rates than in the SM. Clearly the SUSY Higgs search
further enhances the need for the best possible  $m_{\gamma\gamma}$
resolution and $\gamma$-jet rejection.

In extending the SM analysis to the MSSM, one should pay attention to the
applicability of the QCD corrections of ref. [\ref{djouadigg}] to the
gluon-fusion cross-section.
We have checked that in the phenomenologically relevant region, which
corresponds to $h$ or $H$ in the intermediate mass range, and to signal rates
within an order of magnitude from the SM one, the gluon-fusion mechanism is
dominated by the top-quark loop. Since in this region the gluon-fusion
mechanism accounts for $\sim 80 \%$ of the total cross-section, and the
correction is roughly a multiplicative factor 1.5, as a rule of thumb we
can take it into account by multiplying the total cross-section by a factor
$\sim 1.4$.

In fig.~22 we show contour plots in the $(m_A,\tb)$ plane, corresponding
to fixed values of $\sigma \cdot BR (\phi\to\gamma\gamma)$ ($\phi=h,H$).
QCD corrections have been included according to ref. [\ref{djouadigg}].
The region where the rate is large enough to promise a measurable
signal is rather large for $h$, is concentrated in a small strip for $H$,
and is possibly a very small area, just below $m_A = 2 m_t$ and just above
$\tb = 1$, for $A$. In our representative
example [\ref{cseezcap}], we can now evaluate the statistical
significance of the `two-isolated-photons' signal at any point of the
$(m_A,\tb)$ plane, by just combining the information contained in figs.~22,
2, 3, and 21. In the case of $h$ searches, and for $m_h \simgt
90 \gev$, a signal rate beyond 40 fb should give detectable signals.
A signal rate of 30 fb is the borderline of detectability for one year
of running, and signal rates below 20 fb appear extremely difficult to
detect. In the case of $H$, which has higher mass, a signal rate of 20 fb
appears to be the borderline of what can be achieved in one year of running.
In the case of $A$, the interesting mass region is $m_A \sim 2 m_t$:
for $m_A = 250 \gev$, and taking $\sigma \cdot BR (A \to \gamma \gamma)
= 3 \; \fb$ as a plausible discovery limit at the LHC [\ref{cseezcap}],
a signal for $A \to \gamma \gamma$ will be found only if $\tb \simlt 1.5$.

\vskip 0.5cm
{\bf One isolated lepton and two isolated photons. \ }
This signal can come from either $W \phi$ or $\tt \phi$ production.
In the latter case, two or more isolated jets are also produced.
The physics signals from $W \phi$ production are particularly important
at the LHC, and were studied in ref. [\ref{rkzkwjs}]. The
importance of the physics signals from $t \ov{t} \phi$ production
was recently emphasized in ref. [\ref{marcianopaige}].
The production rates, multiplied by the $\phi \to \gamma \gamma$
branching ratio, are shown in fig.~23. We can see that,
similarly to the inclusive $\gamma\gamma$ channel, the rates for
$Wh$, $WH$, $\tt h$, $\tt H$ are always smaller than the  SM value,
which represents the boundary curve in the limit of large $\tb$ and
large (small) $m_A$ for $h$ ($H$).
{}From figs.~23g and 23j one can see that $t \ov{t} A$ production can
give a $l l \gamma$ signal larger than in the SM for small $\tb$ and
near the $m_A = 2 m_t$ threshold, but even in this case the rate appears
to be too small for detection.
We emphasize that the production rates shown in fig.~23
do not include the branching ratios of leptonic $W$ and
semileptonic $t$ decays. If top decays
are as in the SM, one should still include a combinatorial factor of
2, coming from the fact that both top and antitop can decay semileptonically.
On the other hand, in the MSSM there is the possibility of $t \to b H^+$
decays, where the subsequent $H^+$ decay cannot produce a direct lepton
$l=e,\mu$. We shall take this possibility into account in the following,
but its impact on the detectability of the $l \gamma \gamma$ signal is rather
small. The only case in which this effect is not completely negligible is
for $H$, when $m_A \simlt 100 \gev$ and $\tb \simlt 4$ or $\tb \simgt 10$,
in which case the $t \to b H^+$ branching ratio can play a role.

{}From parton-level simulations [\ref{cseez},\ref{kuntrocs},\ref{mangano}],
for a SM Higgs of about 100 GeV, one typically obtains $\sim (12+15)$ and
$\sim (3+11)$ $l \gamma \gamma$ signal events at the LHC and SSC,
respectively. Here we assumed $10^5$ ${\rm pb}^{-1}$ of integrated luminosity
for the LHC and $10^4$ ${\rm pb}^{-1}$ for the SSC. The quoted numbers
separately show the contributions from $W \varphi$ and $t \ov{t} \varphi$
production. Furthermore, they include losses due to acceptances ($\sim
30 \%$), and lepton and photon detection efficiencies [$\epsilon \sim
(0.9)^3$]. The total background is roughly 20--30$\%$ of the signal and
is dominated by the irreducible $W\gamma\gamma$ and $\tt\gamma\gamma$
contributions. There are many different contributions to the reducible
background ($b \ov{b} g, b \ov{b} \gamma, b \ov{b} \gamma \gamma,
W j \gamma, \ldots$). Parton-level simulations indicate that they can
be suppressed well below the irreducible background, provided one assumes,
as for the inclusive $\gamma \gamma$ case, excellent detector performances:
a $\gamma$-jet rejection factor $\simgt 3 \times 10^3$ and a suppression
factor $\simgt 7$ for the leptons from $b$-decays after isolation cuts.

Clearly, there is very little margin (a factor of 2?) to be sensitive to
signal rates smaller than in the SM. In fig.~24, we show contour plots
corresponding to fixed values for the quantity
\be
\label{ldef}
\begin{array}{l}
L_{\phi} \equiv \left[ \sigma \cdot BR \left( l \gamma \gamma \right)
\right]_{\phi}
\\
\\
\phantom{L_{\phi}}
= \bigl[ 2 \sigma (\tt \phi) \cdot BR ( t \to W b)  + \sigma ( W \phi) \bigr]
\cdot BR(\phi\to\gamma\gamma) \cdot BR (W \to l \nu) \, ,
\end{array}
\ee
for the same choice of $m_t$ and $\msq$ as in fig.~15, and for LHC and SSC
energies. In eq. (\ref{ldef}), $l=e,\mu$ and we have not considered the
strongly suppressed possibility
of getting a light charged lepton from both top and antitop.

\vskip 0.5cm
{\bf The four-lepton channel. \ }
The channel $\varphi \to Z^* Z^* \to l^+ l^- l^+ l^-$ ($l=e,\mu$) gives the
so-called `gold-plated' signal for the SM Higgs in the mass range
$m_{\varphi} =$ 130--800 GeV. Below
$m_{\varphi} \sim 130 \gev$, both the total rate and the acceptance decrease
very rapidly, leading to too small a signal for detection. For all three
neutral Higgs bosons of the MSSM, the rates in this channnel are always smaller
than in the SM. In the case of $A$, there is no $AZZ$ coupling at tree level,
and loop corrections cannot generate measurable rates in the four-lepton
channel. As for $h$, if $m_t \simlt 180 \gev$ and $\msq \simlt 1 \tev$,
one can see from fig.~1 that $m_h \simlt 130 \gev$. Therefore, the
$h \to Z^* Z^* \to 4 l$ signal does not have chances of detection at the
LHC and SSC, unless one chooses extremely high values for $m_t$ and
$\msq$ or one has superb resolution and acceptance for leptons.
The situation is somewhat better in the case of $H$, despite the strong
suppression with respect to the SM, due to the competition with the
$hh, b \ov{b}, t \ov{t}$ channels, as discussed in sect.~3.

In fig.~25, we show signal rates for the SM Higgs boson and for $H$, for
the same choices of parameters as in fig.~15. The threshold effects and
the suppression for large values of $\tb$ are clearly visible.

The LHC and SSC discovery potential can be estimated by using the
results of simulations carried out for the SM
[\ref{froid},\ref{nisati},\ref{dellanegra},\ref{evian},\ref{sdcgem}],
taking also into account that $\Gamma_H <
2 \gev$ all over the mass region of interest, $m_H \simlt 2 m_t$.
Assuming excellent lepton momentum resolution, in the
mass range $2 m_Z < m_H < 2 m_t$ a signal rate $\sim 20$ smaller than in
the SM could still lead to a detectable signal. In fig.~26, we show
contour plots in the $(m_A,\tb)$ plane, corresponding to fixed values of
$\sigma \cdot BR ( H \to 4  l)$. QCD corrections have been included according
to ref. [\ref{djouadigg}]. In view of the strong sensitivity to the value
of $m_t$, we show contours for $m_t=120,140,160 \gev$, for LHC and SSC
energies, and for $\msq=1 \tev$. The two almost vertical dashed lines
correspond to $m_H = 2 m_Z$ and to $m_H = 2 m_t$. For $m_H > 2 m_Z$,
a detectable signal could be obtained up to $\tb \sim 5$. Notice that
the experimental acceptances change with $m_H$; in particular, in the region
$m_H < 2 m_Z$ they fastly decrease with decreasing $m_H$: for a realistic
assessment of the discovery limits in this mass region, one should take this
and other effects into account. Anyway, the prospects for detection for
$m_H < 2 m_Z$ do not look good if $m_t \simlt 150 \gev$ and $\msq \simlt
1 \tev$.

\vskip 0.5cm
{\bf The $\taptam$ channel. \ }
For the SM Higgs boson, the $\taptam$ decay channel has been
found hopelessly  difficult for discovery [\ref{dilella},\ref{kbos}],
since this channel has bad mass resolution and overwhelmingly large
background coming from the production of $\tt$, $WW +$ jets,
Drell-Yan pairs, $Z + $ jets, $\bb + $ jets, \ldots .
The bad resolution is due to the fact that the tau-decay products
always include one or more neutrinos, which carry away energy;
therefore one cannot reconstruct the signal as a resonance peak.
The situation is improved if the Higgs is produced with large transverse
momentum that is balanced by a jet [\ref{kellis}].
In this case one can use the approximation
\be
\label{taumiss}
 p_{\nu}^{(1)}{\vec{p}_{l,T}^{\; (1)}\over E_{l}^{(1)}}  +
 p_{\nu}^{(2)}{\vec{p}_{l,T}^{\; (2)}\over E_{l}^{(2)}}  =
 \vec{p}^{\rm \;  miss}_{\rm T}
\ee
to reconstruct the transverse momenta of the neutrinos and hence
the invariant mass of the tau pair. In the above equation,
$ p_{\nu}^{(i)}$ denotes the total transverse momentum of
the neutrinos coming from the decay of $\tau^{(i)}$, $i=1,2$,
while $\vec{p}_{l,T}^{\; (i)}$ and $E_{l}^{(i)}$ denote the lepton momenta
and energies, respectively. It was shown in ref. [\ref{dilella}]
that, in the mass range $m_{\phi}=$ 70--140 GeV,
a mass resolution of $\sim$13--17$\%$ can be achieved.
This method can also be used for the hadronic decay modes,
taking advantage of the fact that the rate is higher by a factor
of $\sim 3.5$. When a tau decays hadronically,
the hadrons have very low multiplicity and invariant mass,
and these properties might be used to recognize the `$\tau$-jet'
[\ref{ua1}].
There is a price for the better mass resolution.
Tagging on a large-$p_{\rm T}$ jet can reduce the rate
by an order of magnitude. Furthermore, at $10^5 \, {\rm pb}^{-1} / {\rm year}$
luminosity, the presence of pile-up deteriorates significantly the
measurement of $\vec{p}_T^{\rm \;\, miss}$, and
therefore the $\taptam$ signal can only be studied
with this method at lower luminosities, $\sim 10^4 \, {\rm pb}^{-1} / {\rm
year}$~\footnote{Alternatively, at high luminosity one may try to just search
for an excess of events in the $e^{\pm} \mu^{\mp}$ or $l^{\pm} +$
`$\tau$-jet' channels.}. While these difficulties appear prohibitive
in the case of the SM, the
situation is not entirely negative in the MSSM.

In fig.~27, we display signal rates for $\phi \to \taptam$ production
($\phi=h,H,A$), for the same parameter choices as in fig.~15, together
with the SM values. We can see that for large values of $\tb$  the
production rates can become much larger than in the SM. In the case of
$h$ production, for $\tb=30$ the enhancement can be more than one order
of magnitude, and increases with decreasing values of $m_h$ (see figs.~27a
and 27d). Huge enhancements can be obtained also for $H$ and $A$, thanks to
the properties of the $\taptam$ branching ratios discussed in sect.~4.
Note for example that, at the LHC, for $m_H,m_A \sim 500 \gev$ and $\tb \simgt
10$, we get  $\sigma \cdot BR(H,A \to \taptam) \geq 20 \; \pb$, while for
$m_A,m_H \sim 120 \gev$ and $\tb \simgt 30$, we obtain $\sigma \cdot BR(H,A
\to \taptam) \geq 30 \; \pb$. The rates for the SSC are rescaled by the
factor already discussed in the previous section.

In order to assess, for any given mass, the cross-section values above
which one obtains a measurable signal over the large background, detailed
simulations are needed. Preliminary studies have been reported for the
lepton channel $e^{\pm} \mu^{\mp}$ in ref. [\ref{dilella}] and for the
mixed channel $l^{\pm} +$ `$\tau$-jet' in ref. [\ref{pausstau}]. In the second
case, the difficulty of recognizing a `$\tau$-jet' may be compensated by the
higher rate of this channel. The preliminary analysis of ref. [\ref{pausstau}]
finds for the LHC sensitivity to values of $\sigma \cdot BR(\phi\to\taptam)$
down to $\sim 10 \; \pb$ in the low-mass region $m_{\phi} \sim 100 \gev$
and $\sim 1 \; \pb$ in the high-mass region $m_{\phi} \sim 400 \gev$.
This result cannot be easily rescaled to the SSC case, since a large
mass interval is involved and the SSC luminosity gives more favourable
experimental conditions for the srudy of this channel.

In fig.~28, we show contour plots corresponding to
fixed values of $\sigma \cdot BR(\phi\to\taptam)$,
for the same values of $m_t$ and $\msq$ as in fig.~15.

\subsection{ Charged Higgs boson }
We now move to the discussion of possible physics signals associated
with the charged Higgs boson. The phenomenology of the charged Higgs
boson at hadron colliders was previously discussed in refs. [\ref{lhcch}].
The benchmark mass value for charged-Higgs-boson searches at the LHC and
SSC is $\mch = m_t - m_b$. For lower values of $\mch$, the dominant
production mechanism at large hadron colliders is $gg \to t \ov{t}$, followed
by $t \to H^+ b$. For higher values of $\mch$, the dominant production
mechanism is $g b \to t H^+$. As far as detectable signals are concerned,
this last case appears hopeless, in view of the suppressed cross-section
and of the large backgrounds coming from QCD subprocesses. The first
case appears instead to be experimentally viable over a non-negligible
region of parameter space. Given the known $t \ov{t}$ production cross-section,
one can compute the $t \to b H^+$ branching ratio according to well-known
formulae, parametrizing again the leading QCD corrections by running
masses evaluated at a scale $Q \sim \mch$. The charged Higgs branching ratios
were discussed in the previous section, where it was found that the
$\tau^+ \nu_{\tau}$ mode dominates in the mass range under consideration.
The experimental signal of a charged Higgs would then be a violation
of lepton universality in semileptonic top decays. As a convenient indicator,
one can consider the ratio
\be
\label{rfelc}
R =
{
BR(t \to \tau^+ \nu_{\tau} b)
\over
BR(t \to \mu^+ \nu_{\mu} b)
}
\equiv
1 + \Delta R \, ,
\ee
with
\be
\label{deltar}
\Delta R =
{
BR(t \to H^+ b) \cdot BR ( H^+ \to \tau^+ \nu_{\tau})
\over
BR(t \to W^+ b) \cdot BR ( W^+ \to \mu^+ \nu_{\mu})
}
{}.
\ee
Preliminary investigations [\ref{lhcch}] show that the experimental
sensitivity could reach $\Delta R \simgt 0.15$ at the LHC. At the
SSC the increased $t \ov{t}$ production cross-section is likely to
give better sensitivity. In fig.~29, we display
contour lines of $\Delta R$ in the $(m_A,\tb)$ plane, for the three
representative values $m_t = 120, \, 140, \,
160 \gev$. The dashed lines denote
the kinematical limit $\mch = m_t - m_b$. One can see that the most
difficult values of $\tb$ are those between 3 and 10, and that the process
under consideration could give access to values of $m_A$ as high as 80--120
GeV for top-quark masses in the range 120--160 GeV.

\section{Conclusions and outlook}

In this paper we carried out a systematic analysis of the possible
physics signals of the MSSM Higgs sector at the LHC and SSC,
assuming that the supersymmetric (R-odd) particles are heavy enough
not to affect significantly the production cross-sections and the
branching ratios of the MSSM Higgs particles. As independent
parameters in the Higgs sector, we chose $m_A$ and $\tb$, and we
considered the theoretically motivated region of the parameter space
$$
0 \le m_A \le 500 \gev,
\;\;\;\;
1 \le \tan \beta \le 50 \, .
$$
We assumed $\msq=1 \tev$ and negligible mixing in the squark sector.
We included
the most important radiative corrections to the Higgs masses $m_h$, $m_H$,
$m_{H^{\pm}}$, and to the Higgs couplings to fermions and vector bosons.
We also included the most important radiative corrections to the three-point
couplings of the neutral Higgs bosons.

We estimated the discovery potential of LEP I and LEP II, and we carried
out detailed cross-section calculations for the LHC and SSC. We singled
out four classes of final states ($\gamma\gamma$, $l^{\pm}\gamma\gamma$,
$l^+l^-l^+l^-$, $\taptam$) which could provide significant signals for
neutral Higgs bosons at the LHC and SSC, and we also examined possible
signals of charged Higgs bosons in top decays.

We calculated all the relevant branching ratios, and the cross-sections
for all the relevant production mechanisms.
We presented our results with the help of branching-ratio curves,
cross-section curves, signal-rate curves and contour plots in the
$(m_A,\tb)$ plane. We did not perform new background
studies, but we pointed out that, using the results of our calculations
and of the existing simulations carried out for the SM Higgs, supplemented
by estimates of the acceptances and efficiencies of typical experiments,
in many cases one can draw conclusions concerning the discovery range. In
some cases, as for the $\taptam$ channel, further simulation work appears
to be needed in order to reach firm conclusions. Nevertheless, some
preliminary conclusions can already be drawn and will now be summarized.

At large hadron colliders, the MSSM Higgs search is, in general, more
difficult than the SM Higgs search.
This is due to the fact that, in a large region of the parameter
space, at least one of the MSSM neutral Higgs bosons
is in the intermediate mass region, $80 \gev \simlt m_{\phi} \simlt 140 \gev$,
but with rates in the $\gamma\gamma$ and $l\gamma\gamma$ channels which can
be significantly suppressed with respect to the SM case. Similarly,
neutral Higgs bosons with $m_{\phi} \simgt 130 \gev$ have typically
strongly suppressed rates in the $l^+l^-l^+l^-$ channel. On the contrary,
in the MSSM, for rather large values of $\tb$, one can obtain a much
larger signal rate in the $\taptam$ channel than in the SM.
Finally, $t \to b H^+$ decays,
followed by $H^+ \to \tau^+ \nu_{\tau}$, can give detectable signals
only in a rather restricted region of the parameter space.

As an example, we now try to assess the discovery potential of the
different channels  for the representative parameter choice $m_t =
140 \gev$, $\msq = 1 \tev$, working as usual in the $(m_A,\tb)$ plane.

To begin with, we recall the expectations for LEP II.
The size of the LEP II discovery region depends rather strongly on
$m_t$ and $\msq$, and on the assumed energy and luminosity.
For standard machine parameters, LEP II cannot test the
whole parameter space allowed by the present data.
Looking back at fig.~8, one may tentatively say that LEP II will give us
(if no Higgs boson is discovered) lower limits of about $m_A \simgt$
70--100 GeV and $\tb \simgt$ 3--8 for $m_t = 120 \gev$, $\tb \simgt$
1.5--3 for $m_t = 160 \gev$.

We observe that the LHC and SSC will test the MSSM
Higgs sector in a largely complementary region of the ($m_A,\tb$) plane.
A pictorial summary of the discovery potential of the different channels
is presented in fig.~30. We emphasize once again that the final discovery
limits will depend on the machine and detector properties, as well as on
the actual values of the top and the soft supersymmetry-breaking masses.
We therefore drew fig.~30 just for illustrative purposes, to exemplify
a particularly convenient way of considering all the discovery channels
at once.

The $\gamma \gamma$ and $l \gamma \gamma$ channels are important in
approximately the same region of the parameter space, $m_A \gg m_Z$ for
$h$ and $50 \gev \simlt m_A \simlt 100 \gev$ for $H$. Therefore, these
two channels can be experimentally cross-checked one against the other,
reinforcing the significance of a possible signal.
As an optimistic discovery limit for $h$, we show in fig.~30 the contour line
$\sigma \cdot
BR (h \to \gamma \gamma) = 30 \; \fb$ at the LHC,  corresponding to
$m_A \simgt 200 \gev$. This contour line is shown only for $m_h \simgt
80 \gev$. In the region of the parameter space to the right of this line
[indicated  by the labels $h\to\gamma\gamma$ and $l + (h\to\gamma\gamma)$],
it is expected that measurable signals will be found, assuming detector and
machine parameters as discussed in refs. [\ref{evian}]. Approximately the same
contour line is obtained by taking $\sigma \cdot
BR (h \to \gamma \gamma) = 85 \;
\fb$ at the SSC. This indicates that, in the inclusive $\gamma \gamma$ channel,
the discovery range of the LHC and SSC will be the same if the luminosity
at the LHC will be  $\sim 3$ times higher than at the SSC and if the
detectors used at the two machines will have similar efficiencies in
suppressing the backgrounds.
Very similar discovery lines can be drawn by considering  the $l \gamma
\gamma$ channel and taking $\sigma \cdot
BR [l + (h\to\gamma\gamma)] \sim 0.8
\; \fb$ for the LHC and $\sim 4 \; \fb$ for the SSC.

In fig.~30 we also show the contour line for $\sigma
\cdot BR (H \to\gamma\gamma)
= 20 \; \fb$ at the LHC, corresponding to $\sim 55 \; \fb$ at the SSC.
The slightly smaller signal rate was chosen to account for the improved
efficiencies at higher Higgs-mass values. The contour line defines a
narrow strip around $m_A\sim 75\gev$ (shaded in fig.~30), where
the discovery of $H$ is expected to be possible both in the
$\gamma \gamma$ and in the $l \gamma \gamma$ channels [for lack
of space the label $l + (H \to \gamma \gamma)$ has been omitted].

The four-lepton channel is important mainly for $H$, in the mass region
$2 m_Z \simlt m_H \simlt 2 m_t$, which translates into $150 \gev \simlt
m_A \simlt 2 m_t$, and for relatively small $\tb$. As a reference
value for discovery in this mass region, we take $\sigma \cdot
BR (H \to 4 l)
= 1 \; \fb$ for the LHC, which corresponds to $\sigma \cdot
BR (H \to 4 l)
\sim 3 \; \fb$ for the SSC. This contour defines the area in fig.~30
indicated  by the label $H\to 4l$.
In a small part of this area, corresponding to $m_A \sim m_t$ and
$\tb \sim 1$, $A \to \gamma \gamma$ could also give a detectable signal.

In the region of very large $\tb$, and moderately large $m_A$, one
could take advantage of the enhanced production cross-sections and of
the unsuppressed decays into $\taptam$ to obtain a visible signal for
one or more of the MSSM neutral Higgs bosons, and in particular for
$H$ and $A$, whose masses can be significantly larger than 100 GeV.
The simulation work for this process is still at a rather early stage
[\ref{pausstau}], so that no definite conclusion can be drawn yet.
For reference, the dotted line in fig.~30 corresponds to a (somewhat
arbitrary) interpolation of $\sigma
\cdot BR (\phi \to \taptam) \sim 10 \, \pb$ at
$m_{\phi} = 100 \gev$ and $\sigma \cdot
BR (\phi \to \taptam) \sim 1\, \pb$ at
$m_{\phi} = 400 \gev$, for LHC energy and summing over the $\phi=H,A$
channels.

Finally, in the region of parameter space corresponding to $m_A
\simlt 100 \gev$, a violation of lepton universality due to the decay
chain $t \to b H^+ \to b \tau^+ \nu_{\tau}$ could indicate the existence of
the charged Higgs boson. This region is indicated by the label
$H^{\pm}\to \tau\nu$ in fig.~30. Its right border is defined
by the contour line of $R = 1.15$, where $R$ was defined in eq.
(\ref{rfelc}).

By definition, our contour lines do not take into account changes in the
acceptances and efficiencies, which are expected in realistic experimental
conditions, and depend on the Higgs mass and on the detector. We therefore
expect some deformations of our contours once discovery lines are extracted
from realistic experimental simulations [\ref{cseezcap},\ref{pausstau}].

As a last piece of information we also display in fig.~30 the border of the
expected discovery region at LEP II, which depends rather sensitively,
as already discussed, on the assumed values of the machine energy and
luminosity.  We then show two representative lines: the lower dashed
line corresponds to $\sigma (e^+ e^- \to
hZ, HZ, hA, HA) = 0.2 \, {\rm pb}$ at $\sqrt{s} = 175 \gev$,
while the upper dashed line corresponds to $\sigma (e^+ e^- \to
hZ, HZ, hA, HA) = 0.05 \, {\rm pb}$ at $\sqrt{s} = 190 \gev$.

Using the result summarized in fig.~30,
we can draw several important qualitative
conclusions:

\begin{itemize}
\item
The discovery potentials of LEP and the LHC/SSC show a certain
complementarity.
The discovery region at LEP covers all $\tb$ values at small values
of $m_A$, and all $m_A$ values at small values of $\tb$, while at
the LHC/SSC one should be sensitive to the large $\tb$, large
$m_A$ region.
\item
One may ask whether the LHC and SSC, combined with LEP II,
can explore the full parameter space of the MSSM Higgs sector, being
sensitive to at least one signal in each point of the $(m_A,\tb$)
plane, for all plausible values of $m_t$ and $\msq$. At present,
this question cannot  be answered positively.
The union of the regions where one should find signals
at least for one Higgs boson does not cover the whole parameter
space: the discovery region has a hole in the middle of the parameter space.
For our parameter choice, the most difficult region appears
to be the cross-hatched area around $m_A=150 \gev$ and $\tb= 10$.
Therefore we cannot claim yet the existence of
a `no-lose' theorem for the MSSM Higgs search.
\item
One may also ask if there are regions of parameter space where one
can find more than one signal from the MSSM Higgs sector. The
answer is that around $m_A=200 \gev$ and $\tb < 5$ one can
discover $h$ at LEP II and $H$ at the LHC/SSC in the four-lepton
channel. There is a somewhat smaller region above $m_A=200\gev$
where one can also find $h$ in the $\gamma\gamma$ and $l \gamma\gamma$
channels. Furthermore, at high values of $\tb $ ($\simgt 20$) and
at $m_A > 200 \gev$ one may discover $A$ and $H$ in the $\tau\tau$ channel
and $h$ in the $\gamma\gamma$ and $l \gamma\gamma$ channels, although
the separation of $H$ and $A$ appears to be impossible, due to their
almost perfect degeneracy in mass. This part of the parameter space is
inaccessible at LEP II. The discovery region for $H$ in the $\gamma\gamma$
and $l, \gamma\gamma$ channels, corresponding to low values of $m_A$,
largely overlaps with the LEP II discovery region and with the discovery
region related to charged-Higgs production in top decays. In the low $\tb$,
for $80 \gev \simlt m_A \simlt 180 \gev$ and $m_A \simgt 2 m_t$ one has
a signal at LEP II and no signal at the LHC and SSC, since $m_h$
is too small for detection.
\end{itemize}

Finally, we would like to make some comments on the theoretical
uncertainties and on possible future studies.

Our values for the signal rates depend on several phenomenological
input parameters, as the value of the bottom mass, the parton-number
densities and the value of $\alpha_S$. The given cross-sections and
branching ratios will change if the input parameters are varied in
their allowed range. Also, for some
production mechanisms, only the Born cross-sections are known.
We estimate that the theoretical errors of the calculated rates
vary from about 30\%, in the case of the $\gamma\gamma$ channel,
up to about a factor of 2 when the $b \ov{b} \phi$ or $t \ov{t} \phi$
production mechanisms are important.

We did not study all effects associated to variations of the parameters
in the SUSY ($R$-odd) sector.
It would be interesting to examine the case when some of the
Higgs bosons are allowed to decay into $R$-odd SUSY particles,
or the effects of squark mixing.
More importantly, serious simulation work is still needed, in particular
for the $\taptam$ and the $l\gamma\gamma$ channel.

\section*{Note added}
After the completion of most of the work presented in this paper, which
was anticipated in many talks [\ref{talks}], we received a number of papers
[\ref{copies},\ref{barger}] dealing with different subsets of the material
presented here, and reaching similar conclusions. Reference [\ref{barger}]
also contains the generalization of eqs. (\ref{hhh1}--\ref{deltahhh})
to the case of arbitrary mixing in the stop and sbottom mass matrices,
but still neglecting the $D$-term contributions.
\section*{Acknowledgements}
We are grateful to G.~Altarelli for discussions, and for insisting
that we should carry out this study. We also thank A.~Brignole, D.~Denegri,
J.~Ellis, L.~Fayard, D.~Froidevaux, J.-F.~Grivaz, P.~Janot, F.~Pauss,
G.~Ridolfi, C.~Seez, T. Sj\"ostrand, D.~Treille, J.~Virdee and P.~Zerwas for
useful discussions and suggestions.
\newpage
\section*{References}
\begin{enumerate}
\item
\label{LPHEP91}
J. Carter,  M. Davier and  J. Ellis, Rapporteur's talks given at the
LP-HEP '91 Conference, Geneva, 1991, to appear in the Proceedings, and
references therein.
\item
\label{higgs}
M.S. Chanowitz, Ann. Rev. Nucl. Part. Phys. 38 (1988) 323;
\\
M. Sher, Phys. Rep. 179 (1989) 273;
\\
R.N. Cahn, Berkeley preprint LBL-29789 (1990);
\\
G. Altarelli, preprint CERN-TH.6092/91.
\item
\label{hunter}
J.F. Gunion, H.E. Haber, G.L. Kane and S. Dawson, {\em The Higgs
Hunter's Guide} (Addison-Wesley, 1990).
\item
\label{MSSM}
For reviews and references see, e.g.:
\\
H.-P. Nilles, Phys. Rep. 110 (1984) 1;
\\
H.E. Haber and G.L. Kane, Phys. Rep. 117 (1985) 75;
\\
R. Barbieri, Riv. Nuo. Cim. 11 (1988) 1.
\item
\label{susyrad}
R. Barbieri, M. Frigeni, F. Giuliani and H.E. Haber, Nucl. Phys. B341
(1990)  309;
\\
A. Bilal, J. Ellis and G.L. Fogli, Phys. Lett. B246 (1990) 459;
\\
M. Drees and K. Hagiwara, Phys. Rev. D42 (1990) 1709;
\\
M. Boulware and D. Finnell, Phys. Rev. D44 (1991) 2054;
\\
M. Drees, K. Hagiwara and A. Yamada, Durham preprint DTP/91/34;
\\
R. Barbieri, M. Frigeni and F. Caravaglios, Pisa preprint IFUP-TH 48/91.
\item
\label{guido}
G. Altarelli, talk given at the LP-HEP '91 Conference, Geneva, 1991, to
appear in the Proceedings, and references therein.
\item
\label{ross}
G.G. Ross, Rapporteur's talk given at the LP-HEP '91 Conference, Geneva,
1991, to appear in the Proceedings, and references therein.
\item
\label{nmssm}
P. Fayet, Nucl. Phys. B90 (1975) 104,
Phys. Lett. 64B (1976) 159 and 69B (1977) 489;
\\
R.K. Kaul and P. Majumdar, Nucl. Phys. B199 (1982) 36;
\\
R. Barbieri, S. Ferrara and C.A. Savoy, Phys. Lett. B119 (1982) 36;
\\
H.P. Nilles, M. Srednicki and D. Wyler, Phys. Lett. B120 (1983) 346;
\\
J.M. Fr\`ere, D.R.T. Jones and S. Raby, Nucl. Phys. B222 (1983) 11;
\\
J.-P. Derendinger and C. Savoy, Nucl. Phys. B237 (1984) 307;
\\
J. Ellis, J.F. Gunion, H.E. Haber, L. Roszkowski and F. Zwirner,
Phys. Rev. D39 (1989) 844;
\\
U. Ellwanger, preprint HD-THEP-91-21, CERN-TH.6144/91;
\\
P. Bin\'etruy and C.A. Savoy, preprint SPhT/91-143, LPHTE Orsay 91/046;
\\
J.R. Espinosa and M. Quir\'os, Madrid preprint IEM-FT-50/91.
\item
\label{muproblem}
J. Polchinski and L. Susskind, Phys. Rev. D26 (1982) 3661;
\\
J.E. Kim and H.-P. Nilles, Phys. Lett. B138 (1984) 150;
\\
L. Hall, J. Lykken and S. Weinberg, Phys. Rev. D27 (1983) 2359;
\\
G.F. Giudice and A. Masiero, Phys. Lett. 206B (1988) 480;
\\
K. Inoue, M. Kawasaki, M. Yamaguchi and T. Yanagida,
Tohoku Univ. preprint TU-373 (1991);
\\
J.E. Kim and H.-P. Nilles, Phys. Lett. B263 (1991) 79;
\\
E.J. Chun, J.E. Kim and H.-P. Nilles, Seoul preprint SNUTP-91-25.
\item
\label{singlet}
H.-P. Nilles, M. Srednicki and D. Wyler, Phys. Lett. B124 (1983) 337;
\\
A.B. Lahanas, Phys. Lett. B124 (1983) 341;
\\
L. Alvarez-Gaum\'e, J. Polchinski and M.B. Wise, Nucl. Phys. B221 (1983)
495;
\\
A. Sen, Phys. Rev. D30 (1984) 2608 and D32 (1985) 411.
\item
\label{gunion}
J.F.~Gunion and H.E.~Haber, Nucl. Phys. B272 (1986) 1,
B278 (1986) 449, and B307 (1988) 445;
\\
J.F.~Gunion, G. Gamberini and S.F.~Novaes, Phys. Rev. D38 (1988) 3481;
\\
T. Weiler and T.C. Yuan, Nucl. Phys. B318 (1989) 337.
\item
\label{kz}
Z. Kunszt and F. Zwirner, in Proceedings of the Large Hadron Collider
Workshop, Aachen, 1990 (G. Jarlskog and D. Rein eds.),
Vol.~II, p.~578 and Erratum.
\item
\label{pioneer}
Y. Okada, M. Yamaguchi and T. Yanagida,
Prog. Theor. Phys. Lett. 85 (1991) 1;
\\
J. Ellis, G. Ridolfi and F. Zwirner, Phys. Lett. B257 (1991) 83;
\\
H.E. Haber and R. Hempfling, Phys. Rev. Lett. 66 (1991) 1815.
\item
\label{zkwjsaa}
Z. Kunszt and W.J. Stirling, in Proceedings of the Large Hadron Collider
Workshop, Aachen, 1990 (G. Jarlskog and D. Rein eds.),
Vol.~II, p.~428.
\item
\label{froid}
D. Froidevaux, in Proceedings of the Large Hadron Collider
Workshop, Aachen, 1990 (G. Jarlskog and D. Rein eds.),
Vol.~II, p.~444.
\item
\label{evian}
ASCOT, CMS, EAGLE and L3/LHC Collaborations, Expressions of Interest
presented at the General Meeting on LHC Physics and Detectors, 5--8
March 1992, Evian-les-Bains, France.
\item
\label{snow}
R. Craven, ed., Proceedings of the 1990 Summer Study on High Energy Physics,
Snowmass, Colo., 1990, to be published, and
references therein.
\item
\label{sdcgem}
Letter of Intent by the Solenoidal Detector Collaboration,
November 1990;
\\
Letter of Intent by the GEM Collaboration, November 1991.
\item
\label{previous}
S.P. Li and M. Sher, Phys. Lett. B140 (1984) 339;
\\
J.F. Gunion and A. Turski, Phys. Rev. D39 (1989) 2701,
D40 (1989) 2325 and D40 (1989) 2333;
\\
M. Berger, Phys. Rev. D41 (1990) 225.
\item
\label{higgscorr}
R. Barbieri, M. Frigeni and M. Caravaglios, Phys. Lett. B258 (1991) 167;
\\
Y. Okada, M. Yamaguchi and T. Yanagida, Phys. Lett. B262 (1991) 54;
\\
A. Yamada, Phys. Lett. B263 (1991) 233;
\\
J.R. Espinosa and M. Quir\'os, Phys. Lett. B266 (1991) 389;
\\
P.H. Chankowski, S. Pokorski and J. Rosiek, Phys. Lett. B274 (1992) 191;
\\
A. Brignole, Padua preprint DFPD/91/TH/28, to appear in Phys. Lett. B;
\\
M. Drees and M.N. Nojiri, preprint KEK-TH-305 (1991);
\\
M.A.~Diaz and H.E.~Haber, Santa Cruz preprint SCIPP-91-14;
\\
A. Brignole, preprint CERN-TH.6366/92, to appear in Phys. Lett. B;
\\
K. Sasaki, M. Carena and C.E.M. Wagner, Munich preprint MPI-Ph/91-109;
\\
S. Kelley, J.L. Lopez, D.V. Nanopolous, H. Pois and K. Yuan,
Texas preprint CTP-TAMU-104/91.
\item
\label{bf}
R. Barbieri and M. Frigeni, Phys. Lett. B258 (1991) 395.
\item
\label{erz3}
J. Ellis, G. Ridolfi and F. Zwirner, Phys. Lett. B262 (1991) 477.
\item
\label{berz}
A. Brignole, J. Ellis, G. Ridolfi and F. Zwirner,
Phys. Lett. B271 (1991) 123.
\item
\label{brignole2}
A. Brignole, unpublished.
\item
\label{lep}
D. Decamp et al. (ALEPH Collaboration), Phys. Lett. B265 (1991) 475;
\\
P. Igo-Kemenes (OPAL Collaboration), L. Barone (L3 Collaboration),
and W. Ruhlmann (DELPHI Collaboration), talks given at the LP-HEP '91
Conference, Geneva, July 1991, to appear in the Proceedings;
\\
M. Davier, Rapporteur's talk at the LP-HEP '91 Conference, Geneva,
July 1991, to appear in the Proceedings, and references therein.
\item
\label{janot}
P. Janot, Orsay preprint LAL 91-61 and talk given at the `Physics at
LEP 200' meeting, Paris, February 1992.
\item
\label{treille}
D. Treille, private communication;
\\
C. Rubbia, Rapporteur's talk given at the LP-HEP '91 Conference, Geneva,
1991, to appear in the Proceedings;
\\
U. Amaldi, plenary talk at the Workshop on Physics and Experiments with
Linear Colliders, Saariselk\"a, September 1991, to appear in the
Proceedings.
\item
\label{gorishnyh}
S.G. Gorishny et al., Mod. Phys. Lett. A5 (1990) 2703, and
references therein.
\item
\label{kniehlbardin}
J. Fleischer and F. Jegerlehner, Phys. Rev. D23 (1981) 2001;
\\
B. Kniehl, Nucl. Phys. B357 (1991) 439;
\\
D.Yu. Bardin, P.Ch. Christova and B.M. Vilenskii, Dubna
preprint JINR-P2-91-140.
\item
\label{djouadigamgam}
A. Djouadi, M. Spira, J.J. van der Bij and  P.M. Zerwas,
Phys. Lett. B257 (1991) 187.
\item
\label{djouadigg}
A. Djouadi, M. Spira and P.M. Zerwas,
Phys. Lett. B264 (1991) 441.
\item
\label{vvzs}
A.I. Vainshtein, M.B. Voloshin, V.I. Zakharov amd M.S. Shifman,
Sov. J. Nucl. Phys. 30 (1979) 711.
\item
\label{loopch}
J.F. Gunion, G.L. Kane and J. Wudka, Nucl. Phys. B299 (1988) 231;
\\
A. M\'endez and A. Pomarol, Nucl. Phys. B349 (1991) 369;
\\
M. Capdequi-Peyran\`ere, H.E. Haber and P. Irulegui,
Phys. Rev. D44 (1991) 231.
\item
\label{qcdch}
M. Drees and K. Hikasa, Phys. Lett. B240 (1990) 455;
\\
A. M\'endez and A. Pomarol, Phys. Lett. B252 (1990) 461;
\\
C.S. Li and R.J. Oakes, Phys. Rev. D43 (1991) 855.
\item
\label{hmrs}
P.N. Harriman, A.D. Martin, R.G. Roberts and W.J. Stirling,
Phys. Rev. { D42} (1990) 798.
\item
\label{georgi}
H.M.~Georgi, S.L.~Glashow, M.E.~Machacek and D.V.~Nanopoulos, Phys. Rev.
Lett. 40 (1978) 692.
\item
\label{wfusion}
M. Chanowitz and M.K. Gaillard, Phys. Lett. 142B (1984) 85 and
Nucl. Phys. B261 (1985) 379;
\\
G.L. Kane, W. Repko and W. Rolnick, Phys. Lett. 148B (1984) 367;
\\
R. Cahn and S. Dawson, Phys. Lett. 136B (1984) 196 and (E)
138B (1984) 464.
\item
\label{zktth}
Z. Kunszt, Nucl. Phys. {B247} (1984) 339;\\
V. Barger, A. Stange and R.J.N. Phillips,  Phys. Rev. D44 (1991) 1987.
\item
\label{sophabgun}
R.M. Barnett,  E. Haber and D.E. Soper,  Nucl. Phys. B306 (1988)
697;\\
D.A. Dicus and S. Willenbrock, Phys. Rev. D39 (1989) 751.
\item
\label{bjorken}
S.L. Glashow, D.V. Nanopoulos and A. Yildiz, Phys. Rev. 18 (1978)
1724.
\item
\label{willen}
 T. Han and S. Willenbrock,
 Phys. Lett. B273 (1991) 167.
\item
\label{cseez}
C. Seez et al., in Proceedings of the Large Hadron Collider
Workshop, Aachen, 1990 (G. Jarlskog and D. Rein eds.),
Vol.~II, p.~474.
\item
\label{cseezcap}
C. Seez and T.S.~Virdee, Imperial College preprint IC/HEP/92-4.
\item
\label{unal}
L. Fayard and G. Unal, EAGLE Internal Note
Physics-NO-001, 1991.
\item
\label{rkzkwjs}
R. Kleiss, Z. Kunszt and W.J. Stirling,
Phys. Lett. {B253} (1991) {269}.
\item
\label{marcianopaige}
W. Marciano and F. Paige,
Phys. Rev. Lett. 66 (1991) 2433;
\\
J.F. Gunion, Phys. Lett. B262 (1991) 510.
\item
\label{kuntrocs}
Z. Kunszt, Z. Trocsanyi and J.W. Stirling, Phys. Lett. B271 (1991)
247;
\\
A.~Ballestrero and E.~Maina, Phys. Lett. B268 (1991) 437.
\item
\label{mangano}
M. Mangano, SDC Collaboration Note SSC-SDC-90-00113 (1990).
\item
\label{nisati}
A. Nisati, in Proceedings of the Large Hadron Collider
Workshop, Aachen, 1990 (G. Jarlskog and D. Rein eds.),
Vol.~II, p.~494.
\item
\label{dellanegra}
M. Della Negra et al., in Proceedings of the Large Hadron Collider
Workshop, Aachen, 1990 (G. Jarlskog and D. Rein eds.),
Vol.~II, p.~509.
\item
\label{dilella}
L. Di Lella, in Proceedings of the Large Hadron Collider
Workshop, Aachen, 1990 (G. Jarlskog and D. Rein eds.),
Vol.~II, p.~530.
\item
\label{kbos}
K. Bos, F. Anselmo and B. van Eijk, in Proceedings of the Large Hadron
Collider Workshop, Aachen, 1990 (G. Jarlskog and D. Rein eds.),
Vol.~II, p.~538.
\item
\label{kellis}
R.K. Ellis, I. Hinchliffe, M. Soldate and J.J. van der Bij,
Nucl. Phys. B297 (1988) 221.
\item
\label{ua1}
C. Albajar et al. (UA1 Collaboration), Phys. Lett. 185B (1987) 233.
\item
\label{pausstau}
F. Pauss, lectures given in the CERN Academic Training Programme,
December 1991, and references therein.
\item
\label{lhcch}
I. Bigi, Y. Dokshitzer, V. Khoze, J. K\"uhn and P. Zerwas,
Phys. Lett. B181 (1986) 157;
\\
V. Barger and R.J.N. Phillips, Phys. Rev. D41 (1990) 884;
\\
A.C. Bawa, C.S. Kim and A.D. Martin, Z. Phys. C47 (1990) 75;
\\
R.M. Godbole and D.P. Roy, Phys. Rev. D43 (1991) 3640;
\\
M. Felcini, in Proceedings of the Large Hadron Collider
Workshop, Aachen, 1990 (G. Jarlskog and D. Rein eds.),
Vol.~II, p.~414;
\\
M. Drees and D.P. Roy, Phys. Lett. B269 (1991) 155;
\\
B.K. Bullock, K. Hagiwara and A.D. Martin, Phys. Rev. Lett.
67 (1991) 3055;
\\
D.P. Roy, preprints CERN-TH.6247/91 and CERN-TH.6274/91.
\item
\label{talks}
Z. Kunszt, talk given to the CMS Collaboration, September 1991;
talk given to the L3/LHC Collaboration, September 1991;
\\
F. Zwirner, talk given to the ASCOT Collaboration, September 1991;
plenary talk at the Workshop on Physics and Experiments with Linear
Colliders, Saariselk\"a, September 1991 and preprint CERN-TH.6357/91;
talk given to the CMS Collaboration, January 1992.
\item
\label{copies}
J.F. Gunion, R. Bork, H.E. Haber and A.Seiden, Davis preprint UCD-91-29,
SCIPP-91/34;
\\
H. Baer, M. Bisset, C. Kao and X. Tata, Florida preprint FSU-HEP-911104,
UH-511-732-91;
\\
J.F. Gunion and L.H. Orr, Davis preprint UCD-91-15;
\\
J.F. Gunion, H.E. Haber and C. Kao, Davis preprint UCD-91-32,
SCIPP-91/44, FSU-HEP-911222.
\item
\label{barger}
V. Barger, M.S. Berger, A.L. Stange and R.J.N. Phillips,
Univ. Wisconsin preprint MAD-PH-680 (1991) (revised).
\end{enumerate}
\newpage
\section*{Figure captions}
\begin{itemize}
\item[Fig.1:]
Contours of $m_h^{max}$ (the maximum value of $m_h$, reached for $m_A \to
\infty$): a)~in the $(m_t,\tan \beta)$ plane, for $\msq = 1 \tev$; b)~in
the $(m_t,\msq)$ plane, for $\tb = m_t / m_b$.
\item[Fig.2:]
Contours of $m_h$ in the $(m_A,\tan \beta)$ plane, for $\msq = 1 \tev$
and a)~$m_t = 120 \gev$, b)~$m_t = 160 \gev$.
\item[Fig.3:]
Contours of $m_H$ in the $(m_A,\tan \beta)$ plane, for $\msq = 1 \tev$
and a)~$m_t = 120 \gev$, b)~$m_t = 160 \gev$.
\item[Fig.4:]
Contours of $m_{H^\pm}$ in the $(m_A,\tan \beta)$ plane, for $\msq =
1 \tev$.
The solid lines correspond to $m_t = 120 \gev$,
the dashed ones to $m_t=160 \gev$.
\item[Fig.5:]
Contours of $\sin^2 \alpha / \cos^2 \beta$ in the $(m_A,\tan \beta)$ plane,
for $\msq = 1 \tev$.
The solid lines correspond to $m_t = 120 \gev$,
the dashed ones to $m_t=160 \gev$.
\item[Fig.6:]
Contours of $\cos^2 \alpha / \sin^2 \beta$ in the $(m_A,\tan \beta)$ plane,
for $\msq = 1 \tev$.
The solid lines correspond to $m_t = 120 \gev$,
the dashed ones to $m_t=160 \gev$.
\item[Fig.7:]
Contours of $\sin^2 (\beta - \alpha)$ in the $(m_A,\tan \beta)$ plane,
for $\msq = 1 \tev$.
The solid lines correspond to $m_t = 120 \gev$,
the dashed ones to $m_t=160 \gev$.
\item[Fig.8:]
Schematic representation of the present LEP I limits and of the future
LEP II sensitivity in the $(m_A,\tan \beta)$ plane, for $\msq = 1 \tev$
and a)~$m_t = 120 \gev$, b)~$m_t = 160 \gev$.
The solid lines correspond to the present LEP I limits.
The dashed lines correspond to $\sigma (e^+ e^- \to
hZ, HZ, hA, HA) = 0.2 \, {\rm pb}$ at $\sqrt{s} = 175 \gev$, which
could be seen as a rather conservative estimate of the LEP II
sensitivity.
The dash-dotted lines correspond to $\sigma (e^+ e^- \to
hZ, HZ, hA, HA) = 0.05 \, {\rm pb}$ at $\sqrt{s} = 190 \gev$, which
could be seen as a rather optimistic estimate of the LEP II
sensitivity.
\item[Fig.9:]
Total widths of the MSSM Higgs bosons, as functions of their respective
masses, for $m_t=140 \gev$, $\msq = 1 \tev$ and $\tb=1.5,3,10,30$: a)~$h$;
b)~$H$; c)~$A$; d)~$H^\pm$.
\item[Fig.10:]
Branching ratios for $h$, as functions of $m_h$, for $m_t=140 \gev$,
$\msq = 1 \tev$ and: a)~$\tb=1.5$; b)~$\tb=30$.
\item[Fig.11:]
Branching ratios for $H$, as functions of $m_H$, for $m_t=140 \gev$,
$\msq = 1 \tev$ and: a)~$\tb=1.5$; b)~$\tb=30$.
\item[Fig.12:]
Branching ratios for $A$, as functions of $m_A$, for $m_t=140 \gev$,
$\msq = 1 \tev$ and: a)~$\tb=1.5$; b)~$\tb=30$.
\item[Fig.13:]
Branching ratios for $h$, as a function of $m_A$, for $m_t=140 \gev$,
$\msq = 1 \tev$ and: a)~$\tb=1.5$; b)~$\tb=30$.
\item[Fig.14:]
Branching ratios for $H^\pm$, as functions of $m_{H^\pm}$, for
$m_t=140 \gev$, $\msq = 1 \tev$ and: a)~$\tb=1.5$; b)~$\tb=30$.
\item[Fig.15:]
Cross-sections for neutral Higgs production, via the gluon-fusion
mechanism, as functions of the corresponding masses and for $m_t=140 \gev$,
$\msq = 1 \tev$, $\tb=1.5,3,10,30$: a)~$h$, LHC; b)~$H$, LHC; c)~$A$,
LHC; d)~$h$, SSC; e)~$H$, SSC; f)~$A$, SSC. QCD corrections are not
included.
\item[Fig.16:]
Cross-sections for $h$ and $H$ production, via the $W$-fusion mechanism, as
functions of the corresponding masses,
for the same parameter choices as in fig.~15:
a)~$h$, LHC; b)~$H$, LHC; c)~$h$, SSC; d)~$H$, SSC.
\item[Fig.17:]
Cross-sections for associated $\bb \phi$ production, as functions of the
corresponding Higgs masses, for
the same parameter choices as in fig.~15:
a)~$h$, LHC; b)~$H$, LHC; c)~$A$,
LHC; d)~$h$, SSC; e)~$H$, SSC; f)~$A$, SSC.
\item[Fig.18:]
Cross-sections for associated $W \phi$ production, as functions of
the corresponding Higgs masses and for
the same parameter choices as in fig.~15:
a)~$h$, LHC; b)~$H$, LHC; c)~$h$, SSC; d)~$H$, SSC.
\item[Fig.19:]
Cross-sections for associated $\tt \phi$ production, as functions of the
corresponding Higgs masses and
for the same parameter choices as in fig.~15:
a)~$h$, LHC; b)~$H$, LHC; c)~$A$,
LHC; d)~$h$, SSC; e)~$H$, SSC; f)~$A$, SSC.
\item[Fig.20:]
Cross-sections times branching ratios for inclusive Higgs production (the
gluon-fusion, $W$-fusion, and $\bb \phi$ contributions are summed) and decay
in the $\gamma\gamma$ channel, for the same parameter choices as in fig.~15:
a)~$h$, LHC; b)~$H$, LHC; c)~$A$, LHC; d)~$h$, SSC; e)~$H$, SSC;
f)~$A$, SSC. For the sake of comparison, the SM values are also indicated.
QCD corrections to the gluon-fusion mechanism are not included.
\item[Fig.21:]
Significance of the inclusive $\phi \to \gamma \gamma$ signal, in the
plane defined by $m_{\phi}$ and $\sigma \cdot BR (\phi \to \gamma \gamma)$,
for the CMS detector proposal at the LHC, with an energy resolution
$\Delta E / E = [2 \% / \sqrt{E(\!\!\gev)}] +0.5 \%$.
The solid lines are contours of
constant $S/\sqrt{B}$, where $S$ is the signal and $B$ is the background.
The dashed line corresponds to the SM Higgs, including QCD corrections
to the gluon-fusion mechanism. Courtesy of C. Seez [\ref{cseezcap}].
\item[Fig.22:]
Contours of constant cross-sections times branching ratios,
in the $(m_A,\tb)$ plane, for the inclusive $\phi\to\gamma\gamma$
channel: a)~$h$, LHC; b)~$H$, LHC; c)~$h$, SSC; d)~$H$, SSC.
The choice of $m_t$ and $\msq$ is the same as in fig.~15, and
QCD corrections to the gluon-fusion mechanism are included.
\item[Fig.23:]
Cross-sections for associated $W\phi$ and $\tt \phi$ production,
times branching ratios for the $\phi\to\gamma\gamma$ channel, for
the same parameter choices as in fig.~15:
a)~$Wh$, LHC; b)~$WH$, LHC; c)~$Wh$, SSC; d)~$WH$, SSC;
e)~$\tt h$, LHC; f)~$\tt H$, LHC; g)~$\tt A$, LHC;
h)~$\tt h$, SSC; i)~$\tt H$, SSC; j)~$\tt A$, SSC.
For the sake of comparison, the SM values are also indicated.
\item[Fig.24:]
Contours of constant  $L_{\phi} =
[ 2 \sigma (\tt \phi) \cdot BR ( t \to W b)  + \sigma ( W \phi) ]
\cdot BR(\phi\to\gamma\gamma) \cdot BR (W \to l \nu)$,
for the same choice of $m_t$ and $\msq$ as in fig.~15:
a)~$h$, LHC; b)~$H$, LHC; c)~$h$, SSC; d)~$H$, SSC.
\item[Fig.25:]
Cross-sections for inclusive $H$ production (the
gluon-fusion, $W$-fusion and $\bb \phi$ contributions are summed) and decay
in the $Z^* Z^* \to 4 l^{\pm}$ channel ($l=e,\mu$),
for the same parameter choices as in fig.~15: a)~LHC; b)~SSC.
For the sake of comparison, the SM values are also indicated.
QCD corrections to the gluon-fusion mechanism are not included.
\item[Fig.26:]
Contours of constant cross-sections times branching ratios for $H \to Z^* Z^*
\to 4 l^{\pm}$, for the same choice of $\msq$ as in fig.~15 and:
a)~$m_t = 140 \gev$, LHC;
b)~$m_t = 140 \gev$, SSC;
c)~$m_t = 120 \gev$, LHC;
d)~$m_t = 120 \gev$, SSC;
e)~$m_t = 160 \gev$, LHC;
f)~$m_t = 160 \gev$, SSC.
QCD corrections to the gluon-fusion mechanism are included.
\item[Fig.27:]
Cross-sections times branching ratios for $\phi \to \tau^+ \tau^-$,
for the same parameter choices as in fig.~15: a)~$h$, LHC; b)~$H$, LHC;
c)~$A$, LHC; d)~$h$, SSC; e)~$H$, SSC; f)~$A$, SSC.
For the sake of comparison, the SM values are also indicated.
\item[Fig.28:]
Contours of constant cross-sections times branching ratios for $\phi \to
\tau^+ \tau^-$, for the same choice of $m_t$ and $\msq$ as in fig.~15:
a)~$h$, LHC; b)~$H$, LHC; c)~$A$, LHC; d)~$h$, SSC; e)~$H$, SSC;
f)~$A$, SSC. The vertical lines in c) and f) correspond to $m_A =
60 \gev$.
\item[Fig.29:]
Contours of constant $\Delta R =[ BR(t \to H^+ b) \cdot BR ( H^+ \to
\tau^+ \nu_{\tau})] / [ BR(t \to W^+ b) \cdot BR ( W^+ \to \mu^+ \nu_{\mu})]$,
for $\msq = 1 \tev$ and: a)~$m_t = 140 \gev$; b)~$m_t = 120 \gev$;
c)~$m_t = 160 \gev$.
\item[Fig.30:]
Pictorial summary of the discovery potential of large hadron colliders
for $\msq = 1 \tev$ and $m_t=140 \gev$.
\end{itemize}
\end{document}